\documentclass[pdflatex,sn-aps,iicol]{sn-jnl}
\usepackage{amsmath,bm}
\usepackage{latexsym}
\usepackage{amssymb}
\usepackage{graphics,epstopdf}
\usepackage{hyperref}
\hypersetup{
	colorlinks = true,
	linkcolor =blue,
	citecolor=blue, 
	urlcolor=blue 
}

\newcommand{\splitatcommas}[1]{%
	\begingroup
	\begingroup\lccode`~=`, \lowercase{\endgroup
		\edef~{\mathchar\the\mathcode`, \penalty0 \noexpand\hspace{0pt plus 1em}}%
	}\mathcode`,="8000 #1%
	\endgroup
}
\usepackage[caption=false]{subfig}
\usepackage[percent]{overpic}
\usepackage{comment}
\usepackage{mathtools}
\usepackage{braket}
\usepackage{geometry}
\newcommand{\mr}[1]{\mathrm{#1}}

\newcommand{\Tr}[2]{\ensuremath{\mathrm{Tr}_{#2}\left[#1\right]}}
\newcommand{\avg}[1]{\ensuremath{\langle #1 \rangle}}
\MakeRobust{\eqref}

\newcommand{\hs}{\hat{\sigma}}
\newcommand{\hsp}{\hat{\sigma}^{+}}
\newcommand{\hsm}{\hat{\sigma}^{-}}
\newcommand{\hsz}{\hat{\sigma}^{z}}
\newcommand{\Hop}{\hat{H}}

\newcommand{\modred}[1]{{\color{black}#1}}

\newcommand{\Bop}{\hat{a}_\mathrm{B}}
\newcommand{\Bopd}{\hat{a}^{\dagger}_\mathrm{B}}

\newcommand{\Lop}{\hat{L}}

\newcommand{\hrho}{\hat{\rho}}
\newcommand{\Uop}{\hat{U}}

\newcommand{\ketbra}[2]{\ensuremath{\vert #1 \rangle \langle #2 \vert}}

\newcommand{\commu}[2]{\ensuremath{\left[#1,#2\right]}}

\newcommand{\mrc}{\mathrm{C}}
\newcommand{\mrb}{\mathrm{B}}

\newcommand{\aop}{\hat{a}}
\newcommand{\adop}{\hat{a}^{\dagger}}
\usepackage[normalem]{ulem}
\usepackage{soul,xcolor}

\begin{document}
	\title{Dephasing Enabled Fast Charging of Quantum Batteries}
	\author[1]{\fnm{Rahul} \sur{Shastri}}
	\author[2,3]{\fnm{Chao} \sur{Jiang}}
	\author[2]{Guo-Hua Xu}
	\author*[1]{B. Prasanna Venkatesh}\email{prasanna.b@iitgn.ac.in}
	\author*[2,4]{Gentaro Watanabe}\email{gentaro@zju.edu.cn}
	\affil[1]{Indian Institute of Technology Gandhinagar, Palaj, Gujarat 382055, India}
	\affil[2]{School of Physics and Zhejiang Institute of Modern Physics, Zhejiang University, Hangzhou, Zhejiang 310027, China}
	\affil[3]{Graduate School of China Academy of Engineering Physics, Beijing 100193, China}
	\affil[4]{Zhejiang Province Key Laboratory of Quantum Technology and Device, Zhejiang University, Hangzhou, Zhejiang 310027, China}
	
	\abstract{ We propose and analyze a universal method to obtain fast charging of a quantum battery by a driven charger system using controlled, pure dephasing of the charger. While the battery displays coherent underdamped oscillations of energy for weak charger dephasing, the quantum Zeno freezing of the charger energy at high dephasing suppresses the rate of transfer of energy to the battery. Choosing an optimum dephasing rate between the regimes leads to a fast charging of the battery. We illustrate our results with the charger and battery modeled by either two-level systems or harmonic oscillators. Apart from the fast charging, the dephasing also renders the charging performance more robust to detuning between the charger, drive, and battery frequencies for the two-level systems case.
	}
	
	\maketitle

    \noindent \section*{Introduction}
	Quantum thermodynamics lies at the intersection of fundamental and applied aspects of quantum information science and technology. This is evident from the problem statements in the field which are typically structured around microscopic versions of macroscopic machines like heat engines, refrigerators, batteries, etc. While the practical utility of such quantum thermal machines is still nascent, the fundamental aspects uncovered by studying them are expected to influence the design and optimization of current and future quantum devices \cite{Quach2023}. In this context, a central goal of the field has been to uncover phenomena that are unique or advantageous in quantum thermal machines such as heat engines, refrigerators, and batteries with no simple counterpart in the classical world \cite{Bhattacharjee2021,myers2022quantum,Quach2023}. In the specific case of quantum batteries (QBs), which are quantum systems that store energy and are charged by direct parametric driving or via ancillary quantum \emph{charger} systems, significant effort has been devoted to identifying situations where figures of merit such as the total energy and ergotropy that can be stored, charging and discharging time, and charging power are optimized \cite{Bhattacharjee2021,Campaioli2018,Andolina2018,GarciaPintos2020,Chen2022,Dou2022,Gyhm2023,Campaioli2023,Hovhannisyan2013,Gallacher2015,Campaioli2017,Ferraro2018,Andolina2019,Ito2020,Rossini2020,Julia-Farre2020, Watanabe2020,Gyhm2022,Shaghaghi2022,Yan2023,Yang2023a,Gyhm2024}.

	
	While early studies of quantum batteries modeled the charging and discharging processes using closed unitary dynamics, there has been a concerted effort recently to extend this paradigm by including dissipative effects \cite{Liu2019,Pirmoradian2019,Barra2019,Farina2019, Quach2020,Tabesh2020,Kamin2020,Mitchison2021chargingquantum,Xu2021,Santos2021,Ghosh2021,Quach2022,Mayo2022,Rodriguez2022,Shaghaghi2023,Yang2023b,Rodriguez2023,Dou2023,Ahmadi2024,gangwar2024coherently}. One motivation to include dissipation stems from the recognition that all realistic quantum battery systems will be subject to interactions with their environment and it is imperative to assess and, if possible, mitigate the negative impact of the resulting dissipation on the battery's performance \cite{Pirmoradian2019,Barra2019,Farina2019,Tabesh2020,Santos2021,Rodriguez2023,gangwar2024coherently}. In contrast, viewing dissipation as a resource, recent works \cite{Liu2019,Quach2020,Gherardini2020,Xu2021,Ghosh2021,Quach2022,Mayo2022,Rodriguez2022,Yang2023b,Dou2023,Ahmadi2024} have highlighted the possibility of charging advantages for dissipative QBs under restricted settings such as specific models of dissipation \cite{Xu2021,Mayo2022}, collective effects \cite{Quach2020,Quach2022,Mayo2022,Dou2023}, control schemes \cite{Mitchison2021chargingquantum,Rodriguez2022}, or particular choices of battery systems \cite{Ghosh2021,Yang2023b}. Nonetheless a simple strategy to obtain fast and stable charging applicable to a wide variety of systems is missing. \modred{Exploring such strategies, especially regarding the limits on the charging time, can also draw from and provide connections to fundamental aspects of quantum dynamics like the quantum speed limit \cite{QSLRef_Deffner_2017,QSLRef_Funo,QSLRef_Pati2}.}
	\begin{figure}
		\centering
		\includegraphics[width=0.8\linewidth]{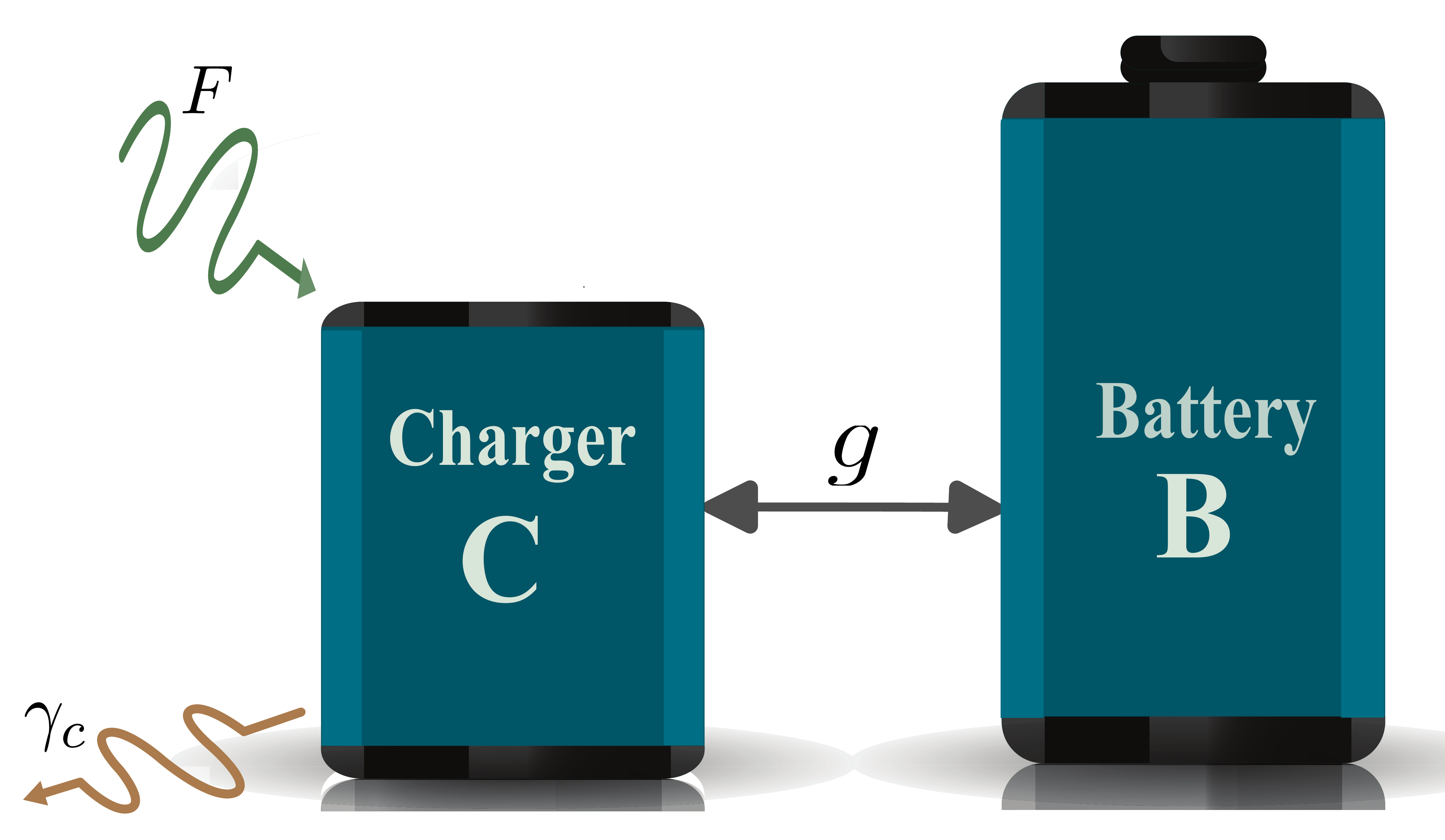}
		\caption{\textbf{Schematic of the charger-battery setup.} Schematic of the setup of a quantum battery B coupled to a quantum charger system C with a coupling constant $g$. The charger is driven at a rate $F$ and additionally subject to dephasing at the rate $\gamma_\mrc$.}
		\label{fig:schematic}
	\end{figure}
	
	Here, we address this challenge by proposing a universal method to obtain fast charging of a quantum battery connected to a driven quantum charger system (shown schematically in Fig.~\ref{fig:schematic}) by controlled pure dephasing of the charger. We note that pure dephasing is one of the fundamental decoherence channels for open quantum systems \cite{nielsen2010quantum} and involves the damping of off-diagonal elements of the density matrix. Pure dephasing is principally generated by the coupling of an operator that commutes with the system Hamiltonian, often taken to be the Hamiltonian itself, with a noisy environment \cite{Skinner86,Lidar2001,Albash2015}. At small values of dephasing, as long as the charger-battery system is initialized in any state apart from the eigenstates of the total Hamiltonian, we expect coherent underdamped oscillation of the battery's energy. In contrast, viewing the pure dephasing process as a continuous measurement of the charger's energy by the environment, strong dephasing naturally leads to a quantum Zeno freezing of the charger's energy and consequent suppression of the rate of charging of the battery. Thus for any charger and battery system, at an in-between moderate value of dephasing that provides a balance between the two effects, we expect to see an optimum fast charging of the battery. This is akin to the working of a shock absorber in a car where adding a dissipative element with the appropriate amount of damping ensures a smooth transfer of mechanical impulse. We confirm and illustrate our general strategy of dephasing-enabled fast charging by choosing the battery and charger systems as two-level systems (TLS) here and demonstrate its wider applicability in the \modred{Methods section} and the supplementary material (SM) with quantum harmonic oscillators (HO) and hybrid TLS-HO setups. Moreover, we also find that the dephased charger lends a certain degree of robustness to the charging process when the charger or its driving is detuned with respect to the battery.
	
	\begin{figure}
		\begin{center}
		\begin{overpic}[width=0.8\linewidth]{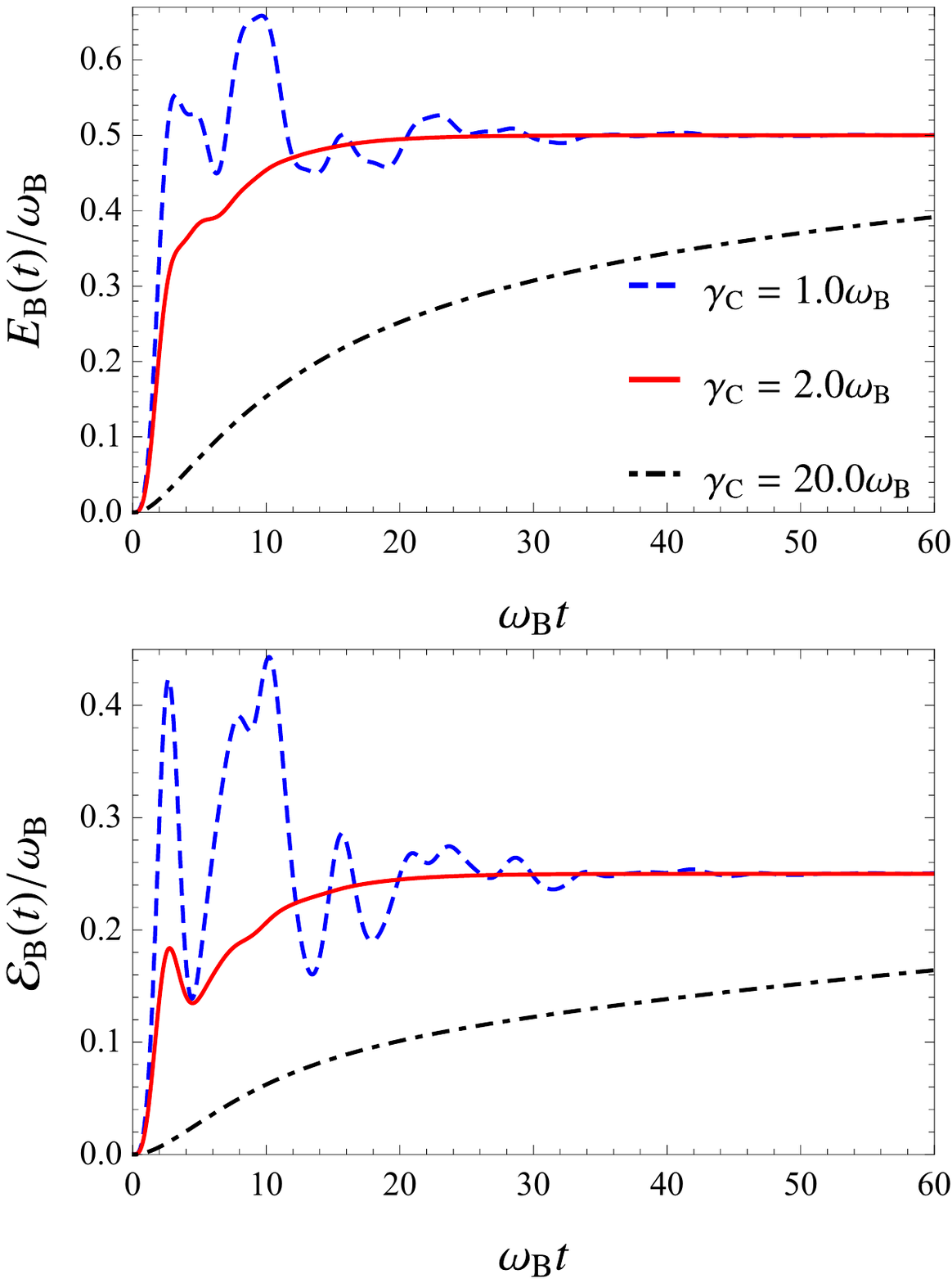}
		    \put(60,90){\textbf{(a)}}
            \put(60,45){\textbf{(b)}}
		\end{overpic}
		\end{center}
		\caption{\textbf{Dynamics of energy and ergotropy of the battery for the two-TLS system.} Time evolution of average energy $E_{\mr{B}}(t)$ [(a)] and ergotropy $\mathcal{E}_{\mr{B}}(t)$ [(b)] of the battery for the two-TLS model for different values of the dephasing rate $\gamma_{\mr{C}}$. Here, as an example, we show the resonant case, i.e. $\omega_\mrc = \omega_\mr{d} = \omega_\mrb$, for the optimal driving with $F=0.5 \omega_\mrb$, and $g = 1.0 \omega_\mrb$ ($F/g=0.5$).
		}
		\label{fig:figTLS2}
	\end{figure}
    \noindent \section*{Results}
    \vspace{-0.35in}
	\noindent \subsubsection*{Setup}
    \noindent We consider a charger-mediated quantum battery setup consisting of a charger system $\mr{C}$, and a quantum battery $\mr{B}$ (see Fig.~\ref{fig:schematic}). The Hamiltonian of the total system has the general form: $\Hop(t)= \Hop_{\mr{C}} +\Hop_{\mr{d}}(t) + \Hop_{\mr{B}} + \Hop_{\mr{CB}}$. Here $\Hop_{\mr{C}}$ and $\Hop_{\mr{B}}$ denote the 
	bare Hamiltonians of the charger and battery system respectively, $\Hop_{\mr{d}}$ provides the coherent driving of the charger system that is the energy source, and $\Hop_{\mr{CB}}$ gives the coupling Hamiltonian between the charger and battery that enables the charging process. In addition to the unitary dynamics generated by $\Hop$, the charger system undergoes a pure dephasing process that will be described via a Gorini-Kossakowski-Sudarshan-Lindblad (GKLS) master equation with a hermitian jump operator $\Lop_\mrc$ that satisfies $[\Lop_\mrc,\Hop_{\mr{C}}] = 0$. Thus the time evolution of the density matrix $\hrho$ describing the charger and battery system is given by
	\begin{align}
		\dot{\hrho} &= -i\commu{\Hop}{\hrho(t)} + \gamma_\mrc \left(\Lop_\mrc \hrho(t) \Lop_\mrc - \frac{\{ \Lop_\mrc^2,\hrho(t)\}}{2} \right)
		\label{eq:GKLSgen},
	\end{align}
	with $\gamma_\mrc$ giving the rate of dephasing and $\{\cdot,\cdot\}$ denoting the anticommutator (we take $\hbar = 1$ throughout). Taking the jump operator as $\Lop_\mrc \propto \Hop_\mrc$, allows the interpretation of the pure dephasing process in Eq.~\eqref{eq:GKLSgen} as resulting from a continuous weak measurement of the energy \cite{wiseman2009quantum} of the charger system. \modred{In Section I of the SM, we provide a detailed derivation of the master equation based on this interpretation.} Starting with the charger and battery system in their respective (free) ground states, the evolution generated by Eq.~\eqref{eq:GKLSgen} leads to an increase in the energy and ergotropy of the battery system that are defined as $E_{\mr{B}} = \Tr{\hrho_{\mr{B}}\Hop_\mrb}{\mr{B}}$ and $\mathcal{E}_{\mr{B}} = E_{\mr{B}} - \min_{\Uop_{\mr{B}}} \Tr{\Uop_{\mr{B}}\hrho_\mrb\Uop_{\mr{B}}^{\dagger}\Hop_{\mr{B}}}{\mr{B}}$, respectively. Here, $\hrho_{\mr{B}} \equiv \Tr{\hrho}{\mr{C}}$ is the reduced density matrix of the battery and the ergotropy is defined by minimization over all possible unitaries $\Uop_{\mr{B}}$ in the battery's Hilbert space. Apart from the values of energy and ergotropy that serve as the standard figures of merit to describe the charging performance, we are specifically interested in the charging time $\tau$ of the battery which we define as
	\begin{align}
		\left \vert \frac{E_\mrb(\tau)-E_\mrb(\infty)}{E_\mrb(0)-E_\mrb(\infty)} \right \vert = e^{-n} \label{eq:taudetexpcondn},
	\end{align}
	with $E_\mrb(\infty)$ denoting the steady state energy of the battery and $n>0$ an integer that we can choose. Note that Eq.~\eqref{eq:taudetexpcondn} can in general have multiple solutions. In our considerations henceforth, we will take $\tau$ as the last root of Eq.~\eqref{eq:taudetexpcondn} such that $|E_\mrb(t)-E_\mrb(\infty)|< e^{-n}|E_\mrb(0)-E_\mrb(\infty)|$ for all $t\geq\tau$. This ensures that we do not underestimate the charging time. \modred{Since achieving the steady state within the GKLS master equation description requires infinite time in principle, our definition of $\tau$ in Eq.~\eqref{eq:taudetexpcondn} is one practical way to estimate the charging time as the time scale over which the transient dynamics of the system settles down and the system's energy approaches the steady state value within some tolerance.}
	\begin{figure*}
		\centering
        \begin{overpic}[width=0.96\linewidth]{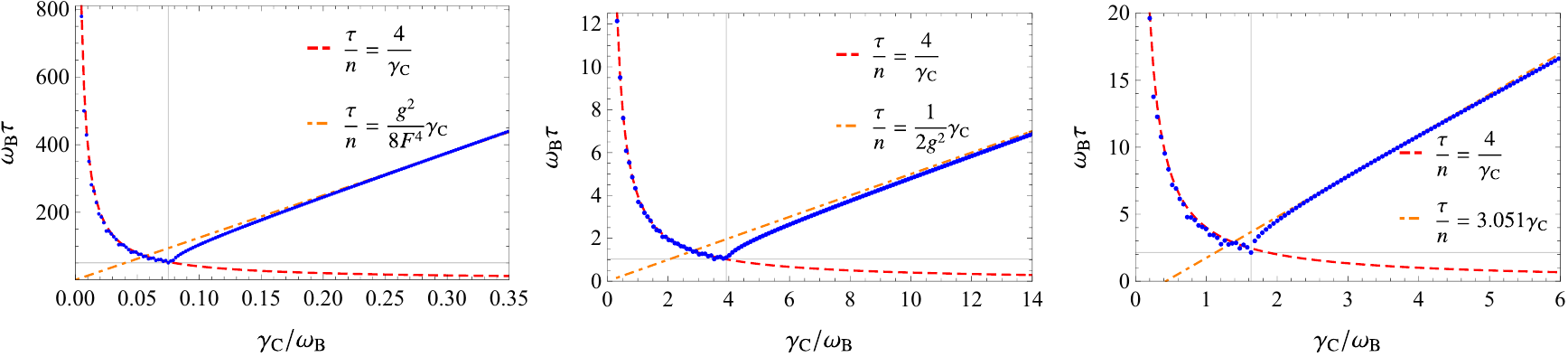}
		    \put(7,18){\textbf{(a)}}
            \put(40,18){\textbf{(b)}}
            \put(75,18){\textbf{(c)}}
		\end{overpic}
		\caption{\textbf{Charging time as a function of dephasing.} Charging time $\tau$ of the battery for the two-TLS model as a function of the charger dephasing rate $\gamma_\mrc$ for different regimes of the driving strength: (a) weak driving $F=0.1 \omega_\mrb$, $g = \omega_\mrb$ ($F/g=0.1$), (b) strong driving $F=10.0 \omega_\mrb$, $g = \omega_\mrb$ ($F/g=10.0$), and (c) optimal driving $F=0.5 \omega_\mrb$, $g = \omega_\mrb$ ($F/g=0.5$). Here, we show the resonant case, i.e., $\omega_\mrc = \omega_\mr{d} = \omega_\mrb$. Gray vertical lines indicate the value, $\gamma_\mrc^\star$, of the optimal dephasing to get the fastest charging time.
		}
		\label{fig:figTLS5}
	\end{figure*}
    To exemplify our results regarding the charging advantages with a dephased charger, we choose the battery and charger as two-level systems with the Hamiltonian
	\begin{align}
		\Hop =&\ \omega_\mrc \hsp_{\mr{C}} \hsm_{\mr{C}}+\omega_\mrb \hsp_{\mr{B}} \hsm_{\mr{B}} + g(\hsp_{\mr{C}} \hsm_{\mr{B}} + \hsm_{\mr{C}} \hsp_{\mr{B}})\nonumber \\
		& + F(\hsm_{\mr{C}} e^{i\omega_{\mr{d}} t} + \hsp_{\mr{C}} e^{-i\omega_{\mr{d}} t}) \label{eq:bareHgen},
	\end{align}
	with $\hsm_{\mr{B,C}}=(\hsp_{\mr{B,C}})^{\dagger}$ representing the Pauli lowering and raising operators, $\omega_\mrb$ ($\omega_\mrc$) giving the frequency of the battery (charger), and $F,\, \omega_{\mr{d}}$ denoting the strength and frequency of the charger drive. When $\omega_\mrb = \omega_\mrc$ (resonance), the battery-charger coupling in Eq.~\eqref{eq:bareHgen} commutes with the bare Hamiltonian of the battery and charger ensuring that there is no energetic cost to switch on/off the interaction in the absence of the driving. In addition, the jump operator that gives the pure dephasing in Eq.~\eqref{eq:GKLSgen} is taken as $\Lop_\mrc = \hsp_{\mr{C}} \hsm_{\mr{C}}$. The energy and ergotropy of the TLS battery can be written as \cite{Farina2019}
	\begin{align}
		E_\mrb &= \frac{\omega_\mrb}{2}\left(\avg{\hs^z_\mrb} + 1\right), \label{eq:TLSEnergy}\\
		\mathcal{E}_\mrb &= \frac{\omega_\mrb}{2}\left(\sqrt{\avg{\hs^z_\mrb}^2 + 4\avg{\hs^+_\mrb}\avg{\hs^-_\mrb}} + \avg{\hs^z_\mrb} \right)
		\label{eq:TLSErgo},
	\end{align} 
	and depend directly on the moments of the TLS operators. Hence, we study the dynamics and properties of energy and ergotropy by solving a closed set of equations for the moments (given in the Methods section) that follows from the master equation~\eqref{eq:GKLSgen}. We begin by considering the resonant case with $\omega_\mrb = \omega_\mrc = \omega_\mr{d}$, where we are able to get exact analytical solutions for the energy and ergotropy (see Methods). 
    \vspace{-0.1in}
	\noindent \subsubsection*{Dephasing Enabled Fast Charging}
	\noindent To illustrate our central result, we plot the time evolution of the average energy $E_{\mr{B}}$ (a) and ergotropy $\mathcal{E}_{\mr{B}}$ (b) in Fig.~\ref{fig:figTLS2} for different values of charger dephasing $\gamma_\mrc$ for a given value of the driving to coupling ratio $F/g = 0.5$ as an example. As evident from the plot, both ergotropy and average energy display underdamped oscillations for very small $\gamma_\mrc$ and slow overdamped behavior at very large $\gamma_\mrc$. Crucially, at an intermediate optimal value of $\gamma_\mrc$ where the dynamics transitions from underdamped to overdamped, the ergotropy and average energy reach their steady values in the shortest time. A qualitative way to understand this behavior is to first note that in the limit where the dephasing is smaller than the interaction $\gamma_\mrc\lesssim g$, the persistent oscillatory exchange of energy between the charger and battery, for our choice of the initial state, dies down slowly leading to slow charging. On the other hand, large dephasing $\gamma_\mrc \gg g$, leads to the charger energy becoming constant due to the quantum Zeno effect. This naturally suppresses the battery charging leading to very slow transfer of energy to the battery. A moderate dephasing rate, as we anticipated, provides a trade-off between the two effects leading to fast charging of the battery. While here we have picked a particular value of $F/g$ for demonstration, this main result holds for any value of $F/g$ as well (see Section II of the SM for details).
	
	Before turning to a detailed analysis of the charging time $\tau$, we first note that for resonant driving the steady state value of the energy is $E_\mrb(t \rightarrow \infty)/\omega_\mrb = \frac{1}{2}$ and the same for the ergotropy takes the form $\mathcal{E}_\mrb(t \rightarrow \infty)/\omega_\mrb = (F/g)/\big(1+4 \frac{F^2}{g^2}\big)$. From this, it is easy to see that the steady state ergotropy is maximized for the optimal value of driving strength $F/g = 0.5$ chosen in Fig. \ref{fig:figTLS2}, similar to previous work reported by Farina \textit{et al}.~\cite{Farina2019}. \modred{As evident from Fig.~\ref{fig:figTLS2}, during the transient dynamics the system's energy can exceed or equal the steady-state values. We find that taking these transient crossings or maxima as charging times is impractical for two reasons. Firstly, the transient nature of the dynamics means that a small perturbation about the identified charging time can lead to much smaller energy values. Secondly, charging the battery to such transient time-instants would require fine tuning of the coupling time between the charger and battery. Moreover, assuming that we can control the coupling time very precisely, quickly removing the coupling between the charger and battery to stop the energy transfer affects the state of the battery, i.e., energy injection/removal by the turning couplings on/off cannot be avoided. Our definition of charging time clearly avoids these issues.} As the first step to systematically study the behaviour of the charging time and the optimal value of dephasing, we calculate $\tau$ by numerically finding roots of Eq.~\eqref{eq:taudetexpcondn}. While the typical choice of $n=1$ in Eq.~\eqref{eq:taudetexpcondn} describes the convergence of the energy to its steady value to one natural logarithm decade, we find that for scenarios with oscillatory terms of higher amplitude, a larger $n$ gives a smoother variation of charging time as other parameters like the dephasing rate are varied. We plot charging time $\tau$, calculated from Eq.~\eqref{eq:taudetexpcondn} with $n=18$, as a function of the dephasing rate $\gamma_{\mr{C}}$ in weak [Fig.~\ref{fig:figTLS5}(a)], strong [Fig.~\ref{fig:figTLS5}(b)], and intermediate [Fig.~\ref{fig:figTLS5}(c)] (which maximizes steady state ergotropy) driving strength regimes, respectively. In all three regimes, we can see that the charging time is minimized at a given optimal value of dephasing, denoted as $\gamma_\mrc^\star$ henceforth, underscoring our central result that moderate dephasing leads to fast charging. Additionally, we note that taking even larger values of $n>18$ in the calculation of $\tau$ will only make the behavior of the charging time in Fig.~\ref{fig:figTLS5} smoother but not affect any of the important qualitative features, especially the value of $\gamma_\mrc^\star$ (see in particular Fig.~\ref{fig:figTLS5}(c) where the scattering of the data points is still discernible).
	
	Exploiting the exact expression for the battery's average energy, we next obtain analytic expressions for the charging time at large ($\gamma_\mrc\gg g$) and small dephasing ($\gamma_\mrc \ll g$)  regimes. This will allow us to also provide explicit analytical estimates for the optimal dephasing rate $\gamma_\mrc^\star$. Leaving the details of the derivation to the Methods section, we note that in the small dephasing limit ($\gamma_{\mrc}\ll g$) the charging time scales as,
	\begin{align}
		\tau \sim \frac{4n}{\gamma_\mrc}  
		\label{eq:taugammaSmall2},
	\end{align}
	irrespective of the values of $F$ and $g$. This is neatly illustrated in Figs.~\ref{fig:figTLS5}(a), \ref{fig:figTLS5}(b), and \ref{fig:figTLS5}(c), where the behavior of the calculated charging time compares well with the red dashed curve representing the scaling given by Eq.~\eqref{eq:taugammaSmall2} at small $\gamma_\mrc$. On the other hand, in the large dephasing limit ($\gamma_{\mrc} \gg g$), we find
	\begin{align}
		\tau &\sim \frac{ng^2}{F^4}\gamma_\mrc \,\, \mathrm{for} \,\, F/g \ll 1,
		\label{eq:tauweakdephasing} \\
		\tau &\sim \frac{n}{2g^2}\gamma_\mrc \,\, \mathrm{for} \,\, F/g \gg 1
		\label{eq:taugammaLarge2}.
	\end{align}
	This large dephasing scaling is clearly illustrated by the orange dash-dotted lines in Fig.~\ref{fig:figTLS5}(a) for the weak driving $F/g\ll 1$ and Fig.~\ref{fig:figTLS5}(b) for the strong driving $F/g\gg 1$. The scaling of charging time linearly with $\gamma_\mrc$ in the large dephasing limit  also holds for intermediate driving with $F\sim g$. We show this in Fig.~\ref{fig:figTLS5}(c) and also determine the pre-factor of the linear scaling using a numerical fit (orange dash-dotted line). The optimum dephasing rate $\gamma_\mrc^\star$ that leads to the fastest charging lies precisely in between the distinct scaling behaviors we have uncovered above for large and small $\gamma_\mrc$. As we show in Section II of the SM, an accurate estimate for the optimum dephasing rate $\gamma_\mrc^\star$ can be obtained from the value of $\gamma_\mrc$ for which the average energy's dynamics shows an underdamped to overdamped transition. From this analysis we find that $\gamma_\mrc^\star \sim 8 F^2/g$ for $F \ll g$ and $\gamma_\mrc^\star \sim 4 g$ for $F \gg g$. We can also arrive approximately at this estimate for $\gamma_\mrc^\star$ by simply equating the charging time in the small $\gamma_\mrc$ limit Eq.~\eqref{eq:taugammaSmall2} to the ones in the large $\gamma_\mrc$ case given in Eqs.~\eqref{eq:tauweakdephasing} and \eqref{eq:taugammaLarge2}. Note that one can also define the fastest charging time $\tau^\star \equiv 4/\gamma_\mrc^\star$ (using Eq.~\eqref{eq:taugammaSmall2} with $n=1$) and it takes the value $\tau^\star \sim g/(2F^2)$ and $\tau^\star \sim 1/g$ for weak and strong driving, respectively.

	\begin{figure}
		\centering
		\begin{overpic}[width=0.8\linewidth]{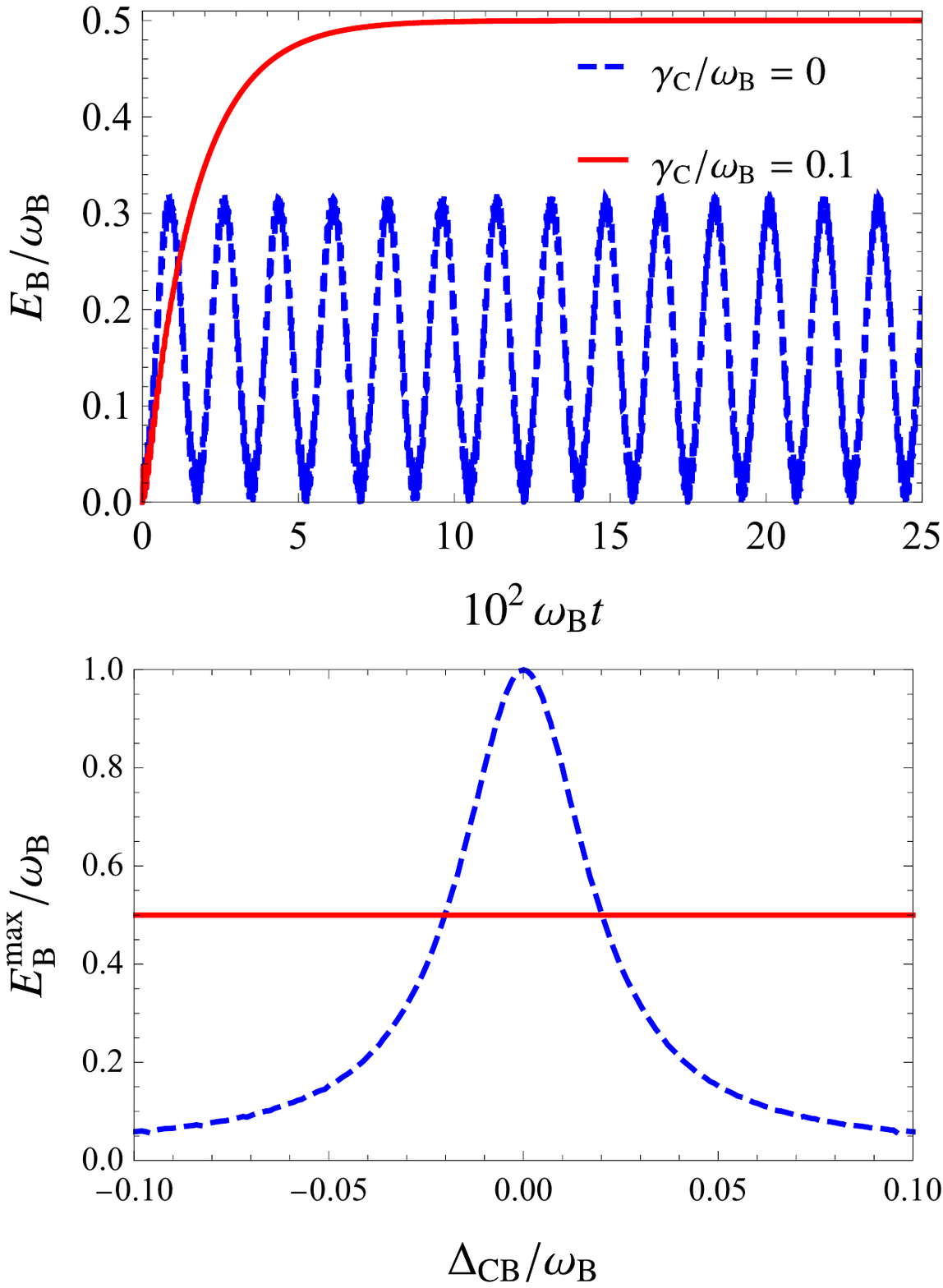}
		    \put(25,90){\textbf{(a)}}
            \put(25,40){\textbf{(b)}}
		\end{overpic} 
		\caption{\textbf{Energy of the battery coupled to a detuned charger.} (a) Transient dynamics of the average energy $E_{\mrb}(t)$ of the battery in the weak driving regime $F/g=0.1$ for the detuned charger-battery case with $\Delta_{\mr{CB}}=0.03 \omega_{\mrb}$. (b) Maximum value of the oscillating energy, $E_\mrb^{\mr{max}}$, in the closed case (blue dashed line) is compared to the case with dephasing (red line) as function of the charger-battery detuning. Other parameter values are $g=\omega_{\mrb}$ and $\omega_{\mr{d}}=\omega_\mrb$.}
		\label{fig:figTLS6}
	\end{figure}
    \vspace{-0.1in}
    \noindent \subsubsection*{Robust Charging against Detuning}
	Let us now consider the charging behavior when the charger and battery are not at resonance. We first note that our central result of fast charging with moderate dephasing holds for such detuned cases as well. In addition, we want to highlight an additional advantage that emerges in comparison to the case without dephasing. Figure~\ref{fig:figTLS6}(a) shows the transient dynamics of the average energy $E_{\mrb}(t)$ of the battery in the weak driving regime $F/g=0.1$ for an exemplary case with the battery and charger detuned and the driving resonant with the battery, i.e., $\omega_\mrb = \omega_\mathrm{d} \neq \omega_\mrc$ and $\Delta_{\mr{CB}}=\omega_\mrc - \omega_\mrb$. While the energy of the battery oscillates without damping for the closed ($\gamma_\mrc = 0$) case, for the non-zero dephasing case the energy attains the stable value $\omega_\mrb/2$ (with $\gamma_\mrc$ chosen to minimize the charging time). Remarkably, this steady state value is larger than the oscillating closed-case values. Moreover, we also find that, in the presence of charger dephasing, the battery ergotropy also attains steady state value larger than the (oscillating) ergotropy for the closed case. In this sense we conclude that dephasing can provide a degree of robustness against detuning in terms of charging performance. This robustness is summarized in Fig.~\ref{fig:figTLS6}(b), where we compare the energy in the closed and dephased cases as a function of the battery-charger detuning $\Delta_{\mr{CB}}$.
    
    \modred{Evidently, in the closed case since the drive-charger detuning $\Delta_{\mathrm{Cd}}$ is taken equal to the battery-charger detuning $\Delta_{\mathrm{CB}}$ (i.e., $\omega_\mr{d} = \omega_\mrb$), the maximum value of the oscillating energy of the battery has a Lorentzian behaviour centered at $\Delta_{\mathrm{CB}}= \Delta_{\mathrm{Cd}}=0$ stemming from the reduced polarization of the charger qubit and consequently suppressed energy transfer to the battery when the driving is detuned. In contrast, as shown in Fig.~\ref{fig:figTLS6}(b), for the dephased charger the steady-state value of the energy is independent of detuning and becomes greater than the closed case for a wide range of detuning values away from resonance $\Delta_{\mathrm{CB}} = \Delta_{\mathrm{Cd}} =0$. Since dephasing can be thought of as arising from the addition of frequency noise to the charger system (see Section I of the SM), the charger has some frequency ‘uncertainty’ or frequency linewidth. This helps it overcome the strict detuning with the battery frequency and enables better transfer of energy than the closed case. Further details, including similar behavior for the ergotropy at weak driving as well as cases with charger driving detuned from the battery are presented in the SM (Section II).}
	\noindent \section*{Discussion}
    \noindent In summary, we have demonstrated the advantages of a dephased charger setup for the TLS battery. Our result provides a strategy for the most advantageous way to charge a TLS battery with a charger. To that end, we have to first take the ratio between the driving and coupling strength $F/g = 0.5$ to obtain the largest steady-state ergotropy of $\mathcal{E}_{\mrb}(t\rightarrow \infty) = 0.25 \omega_\mrb$. To obtain the fastest charging time, notice from Fig.~\ref{fig:figTLS5}(c) that the optimal dephasing for the moderate driving regime of $F/g = 0.5$ is given by $\gamma_\mrc^{\star} \approx 1.15 \omega_\mrb = 2.3 F$. Since the charging time scales inversely with $\gamma_\mrc^{\star}$, we can further get the fastest charging by applying the optimum dephasing $\gamma_\mrc^{\star}$ with the largest allowed value of $F$ in the particular physical realization. \modred{While we have discussed the two-TLS setup extensively here, as shown in the Methods section (see also Sections III and IV of the SM), our central result is readily demonstrable for the two-HO and TLS-HO setups, underscoring its wide applicability.} \modred{In our discussions of battery charging performance, we have focused on the energy and ergotropy as figures of merit. Since we are interested in a dissipative charging scenario, another additional variable of importance is the von Neumann entropy of the battery. We find that in general, for all the different settings we have considered, higher accumulated entropy leads to lower ergotropy as expected. Nonetheless, the steady state ergotropy with dephasing can always be tuned to a significant value by choosing the driving strength $F/g$ appropriately or also by collectively coupling the charger to multiple battery TLSs (see Fig.~S3 of the SM). Additionally, as we discuss in the SM (Section II), for the two-TLS case we can also express the entropy as a simple function of the difference between the energy and ergotropy $(E_\mrb-\mathcal{E}_\mrb)$.}

	Quantum batteries are energy storage devices that are susceptible to environmental dissipation effects. The natural question, then, is whether one can utilize these dissipative processes to enhance the overall performance of quantum batteries. In this article, we have analyzed a general method to leverage the dephasing of the charger to get fast, stable, and robust charging of a quantum battery. Our strategy, illustrated with TLS and HO models for charger and battery, relies on subjecting the charger to an optimal amount of dephasing that provides the appropriate trade-off between coherent oscillatory dynamics of energy expected at a small dephasing rate and the slow exchange of energy expected from the quantum Zeno effect at a large dephasing rate. Since this competition between coherent exchange and dephasing can occur for a wide variety of quantum systems, our main finding that moderately dephased chargers lead to efficient charging should hold universally. Moreover, for the two-TLS charger-battery setup, the dephasing also provides robustness, in terms of charging energy and ergotropy, to detuning the driving from the charger and battery. Finally, we would like to emphasize that while the focus here has been on steady state charging, in the transient regime the battery can attain values of energy and ergotropy greater than their respective steady state values. Such transient oscillations are especially pronounced for small values of $\gamma_\mrc$. Thus, our strategy of moderate dephasing to attain robust and fast charging to a steady can be complemented by other approaches exploiting this transient advantage, albeit with the additional requirement of precise control in switching the charger-battery coupling.
	
Our theoretical findings can be readily verified in state-of-the-art quantum technology platforms. For instance, recently experimental implementations of quantum batteries have been achieved with superconducting qubits \cite{Hu2022}, NMR \cite{Joshi2022}, and organic semiconductor microcavity systems \cite{Quach2022}. In all of these realizations, the battery and charger systems are indeed exposed to an environment that can lead to dissipation and decoherence. Moreover, a key feature needed to implement our central idea, namely the ability to control the dephasing strength, has already been demonstrated in the experimental platforms of superconducting qubits \cite{Berger2015measurement} and NMR systems \cite{Joshi2022}. Finally, while we have considered single-component battery and charger systems here, motivated by the possibility of enhancing the steady-state ergotropy of the battery (see SM), exploring collective battery-charger systems with dephasing can be an interesting future research direction \cite{Campaioli2017,Joshi2022}.
    \noindent \section*{Methods}
    \vspace{-0.35in}
	\noindent \subsubsection*{Energy of the TLS Battery and Charger}
	\noindent The dynamics for the moments of the TLS battery-charger system are conveniently written down by moving to an interaction picture by considering a unitary transformation $\hat{U}_{\mr{CB}}=\exp(-i[\omega_\mrc \hsp_{\mr{C}}\hsm_{\mr{C}}+\omega_\mrb \hsp_{\mr{B}}\hsm_{\mr{B}}]t)$ resulting in the following Hamiltonian
	\begin{align}
		\Hop^\prime =&\  g\left( \hsp_{\mr{C}}\hsm_\mrb e^{i \Delta_{\mr{CB}}t}+ \hsm_\mrc \hsp_\mrb e^{-i \Delta_{\mr{CB}} t} \right) \nonumber\\ 
		&+ F(\hsm_\mrc e^{-i\Delta_{\mr{Cd}} t} + \hsp_\mrc e^{i\Delta_{\mr{Cd}} t}) \label{eq:HintTLS},
	\end{align}
	with the frequency differences $\Delta_{\mr{CB}} = \omega_\mrc-\omega_\mrb$ and $\Delta_{\mr{Cd}} = \omega_{\mr{C}}-\omega_\mr{d}$. Since the jump operator $\Lop_\mrc = \hsp_\mrc \hsm_\mrc$ is invariant under this unitary, the master equation \eqref{eq:GKLSgen} in the interaction picture reads
	\begin{align}
		\frac{d\hrho^\prime(t)}{dt} &= -i\commu{\Hop^{\prime}}{\hrho^\prime(t)} + \frac{\gamma_\mrc}{4} \left(\hsz_\mrc \hrho^\prime(t) \hsz_\mrc - \hrho^{\prime}(t)\right)
		\label{eq:TLSmasterinteraction},
	\end{align}
	with $\hrho^\prime = \hat{U}_{\mr{CB}}^{\dagger} \hrho \hat{U}_{\mr{CB}}$. We can write down a closed set of equations for the moments (first and second) of the TLS operators from the master equation \eqref{eq:TLSmasterinteraction} which read as:
	\begin{align}
		\frac{d}{dt}\avg{\hs^z_{\mr{B}}} =& -4g\mathfrak{Im}\left[e^{i\Delta_\mr{{CB}}t}\avg{\hs^+_{\mr{C}}\hs^-_\mrb}\right], \nonumber\\
		\frac{d}{dt}\avg{\hs^z_{\mr{C}}} =& -4F \mathfrak{Im}\left[ e^{-i\Delta_\mr{{Cd}}t}\avg{\hs^-_{\mr{C}}}\right] \nonumber \\
		&\hspace{2cm} + 4g \mathfrak{Im} \left[ e^{i\Delta_\mr{{CB}}t}\avg{\hs^+_{\mr{C}}\hs^-_{\mr{B}}}\right] \nonumber, \\
		\frac{d}{dt} \avg{\hs^+_{\mr{C}}\hs^-_{\mr{B}}} =& -iFe^{-i\Delta_\mr{{Cd}}t}\avg{\hs^z_{\mr{C}}\hs^-_{\mr{B}}} -\frac{\gamma_{\mr{C}}}{2}\avg{\hs^+_{\mr{C}}\hs^-_{\mr{B}}} \nonumber\\
		& -i\frac{g}{2}e^{-i\Delta_\mr{{CB}}t}\left(\avg{\hs^z_{\mr{C}}} - \avg{\hs^z_{\mr{B}}} \right), \nonumber \\ 
		\frac{d}{dt}\avg{\hs^z_{\mr{C}}\hs^-_{\mr{B}}} =& -2iF\left(e^{i\Delta_\mr{{Cd}}t}\avg{\hs^+_{\mr{C}}\hs^-_{\mr{B}}} -  e^{-i\Delta_\mr{{Cd}}t}\avg{\hs^-_{\mr{C}}\hs^-_{\mr{B}}}\right) \nonumber\\
		&+ ige^{-i\Delta_\mr{{CB}}t}\avg{\hs^-_{\mr{C}}}, \nonumber\\
		\frac{d}{dt}\avg{\hs^-_{\mr{C}}\hs^-_{\mr{B}}} =&\ iFe^{i\Delta_\mr{{Cd}}t}\avg{\hs^z_{\mr{C}}\hs^-_{\mr{B}}} - \frac{\gamma_{\mr{C}}}{2}\avg{\hs^-_{\mr{C}}\hs^-_{\mr{B}}}, \nonumber\\
		\frac{d}{dt}\avg{\hs^-_{\mr{C}}} =&\ iFe^{i\Delta_\mr{{Cd}}t} \avg{\hs^z_{\mr{C}}} + ige^{i\Delta_\mr{{CB}}t}\avg{\hs^z_{\mr{C}}\hs^-_{\mr{B}}} \nonumber\\
		&- \frac{\gamma_{\mr{C}}}{2}\avg{\hs^-_{\mr{C}}},
		\label{twoqubitsEOMs1}
	\end{align}
	and
	\begin{align}
		\frac{d}{dt}\avg{\hs^-_{\mr{B}}} =&\ ige^{-i\Delta_\mr{{CB}}t} \avg{\hs^-_{\mr{C}}\hs^z_{\mr{B}}}, \nonumber\\
		\frac{d}{dt}\avg{\hs^-_{\mr{C}}\hs^z_{\mr{B}}} =&\ iFe^{i\Delta_\mr{{Cd}}t}\avg{\hs^z_{\mr{C}}\hs^z_{\mr{B}}} + ige^{i\Delta_\mr{{CB}}t}\avg{\hs^-_{\mr{B}}} \nonumber\\
		&- \frac{\gamma_{\mr{C}}}{2}\avg{\hs^-_{\mr{C}}\hs^z_{\mr{B}}},  \nonumber\\
		\frac{d}{dt}\avg{\hs^z_{\mr{C}}\hs^z_{\mr{B}}} =& -4F\mathfrak{Im}\left[e^{-i\Delta_\mr{{Cd}}t}\avg{\hs^-_{\mr{C}}\hs^z_{\mr{B}}}\right].
		\label{twoqubitsEOMs3}
	\end{align}
	\modred{Note that the above equations are determined purely by the ratios $F/g$ and $\gamma_\mrc/g$ in the resonant case with $\Delta_{\mr{Cd}}=\Delta_{\mr{CB}}=0$. While we have chosen $g=\omega_{\mrb}$ in the results shown in Figs.~\ref{fig:figTLS2} and \ref{fig:figTLS5}, the results remain exactly the same for even smaller coupling $g$, as long as we make a proportional change in $F$ and $\gamma_\mrc$ keeping the same ratios $F/g$ and $\gamma_\mrc/g$. Thus, we can also understand our results as being calculated for small values of $g/\omega_\mrb$. Note that taking $g/\omega_\mrb$ to smaller values also leads to smaller values of $\gamma_\mrc/\omega_\mrb$ required to obtain the dephasing enabled charging advantage as per the scaling we described. Viewing our results as being in this regime with weak system-bath ($\gamma_\mrc \ll \omega_\mrb$)  and charger-battery ($g \ll \omega_\mrb$) coupling provides additional justification for Eq.~\eqref{eq:GKLSgen} as these are precisely the conditions required in standard derivations of a local GKLS master equation \cite{breuer_theory_2007}.} Choosing the density matrix at initial time as $\hrho^\prime(t=0) = \ketbra{g}{g}_\mrb \otimes \ketbra{g}{g}_\mrc$ with $\ket{g}_\mrb$ ($\ket{g}_\mrc$) denoting the ground state of the battery (charger), the initial conditions for the moments become: 
	\begin{align}
		\avg{\hs^z_{\mr{B}}}(0) &= \avg{\hs^z_{\mr{C}}}(0) = -1, \nonumber \\
		\avg{\hs^+_{\mr{C}}\hs^-_{\mr{B}}}(0) &= \avg{\hs^z_{\mr{C}}\hs^-_{\mr{B}}}(0) = \avg{\hs^-_{\mr{C}}\hs^-_{\mr{B}}}(0) = \avg{\hs^-_{\mr{C}}}(0) = 0, \nonumber\\
		\avg{\hs^-_{\mr{B}}}(0) &= \avg{\hs^-_{\mr{C}}\hs^z_{\mr{B}}}(0) = 0, \nonumber\\
		\avg{\hs^z_{\mr{C}}\hs^z_{\mr{B}}}(0) &= 1.
		\label{twoqubitsEOMsInitialCond}
	\end{align}
	Solving Eqs.~\eqref{twoqubitsEOMs1} and \eqref{twoqubitsEOMs3} with initial conditions Eq.~\eqref{twoqubitsEOMsInitialCond} and using Eq.~\eqref{eq:TLSEnergy}, we can write down the exact expressions for the average energy of the battery $E_{\mrb}(t)$ as
	\begin{align}
		&\frac{E_{\mrb}(t)}{\omega_\mrb} = \frac{1}{2} -\frac{e^{-\frac{\gamma_\mrc}{4}t}}{4\left(1+4\frac{F^2}{g^2}\right)} \Bigg\{\frac{8F^2}{g^2} \chi_t(\gamma_\mrc,g,f_0)  \nonumber \\
		& \left .  +  \left(1+\sqrt{1+\frac{4F^2}{g^2}}\right) \chi_t(\gamma_\mrc,g,f_1) \right . \nonumber \\
		& + \left(1-\sqrt{1+\frac{4F^2}{g^2}}\right) \chi_t(\gamma_\mrc,g,f_2)
		\Bigg\}  \label{eq:EnergyanalytFnTLSres},
	\end{align}
	with 
	\begin{align}
		&\chi_t(\gamma_\mrc,g,f_i) =  \left[\cosh\left(\frac{\sqrt{\gamma_\mrc^2-32f_ig^2}}{4}t\right) \right .  \nonumber \\
		&\left . \left . \hspace{1cm} + \frac{\gamma_\mrc}{\sqrt{\gamma_\mrc^2-32f_ig^2}}   
		\sinh\left(\frac{\sqrt{\gamma_\mrc^2-32f_ig^2}}{4}t\right)\right] \right.
		\label{eq:chifn},
	\end{align}
	$f_0 = 1/2$, $f_1 =\left(1+2F^2/g^2 -\sqrt{1+4F^2/g^2}\right)$, and $f_2 =\left(1+2F^2/g^2 + \sqrt{1+4F^2/g^2}\right)$. In a similar manner we can also write down analytical expressions for the ergotropy (see Section II of the SM) but the analytical expressions for the ergotropy are too cumbersome to present and do not add insight. Nonetheless, we note that results in Fig.~\ref{fig:figTLS2} were produced by evaluating the expressions using Mathematica ~\cite{Mathematica}.
	\noindent \subsubsection*{Analytical Expressions for charging time and optimal dephasing for TLS Battery-Charger system}
	\noindent Exploiting the exact expression for the battery's average energy, we can obtain analytic expressions for the charging time at large ($\gamma_\mrc\gg g$) and small dephasing ($\gamma_\mrc \ll g$)  regimes. This will allow us to support the analytical estimates for the charging time and optimal dephasing rate $\gamma_\mrc^\star$ presented in the main text. Taking the small dephasing limit ($\gamma_{\mrc}\ll g$) in the exact expression of the energy, Eq.~\eqref{eq:taudetexpcondn} that determines the charging time takes the form
	\begin{align}
		\left \vert e^{-\frac{\gamma_{\mrc}}{4}\tau}\phi(F,g,\gamma_\mrc,\tau)\right \vert = e^{-n}
		\label{eq:taugammaSmall1},
	\end{align}
	where the function $\phi(F,g,\gamma_\mrc,t)$ can be determined as follows. In the limit of $\gamma_\mrc \ll g$, the hyperbolic trigonometric functions determining the function $\chi_t(\gamma_\mrc,g,f_i)$ in the energy expression \eqref{eq:EnergyanalytFnTLSres} will become standard trigonometric functions. The small $\gamma_\mrc \ll g$ approximation to $\chi$ as $\tilde{\chi}$, i.e., $\chi_t(\gamma_\mrc,g,f_i)  \stackrel{\gamma_\mrc \ll g}{\approx} \tilde{\chi}_t(\gamma_\mrc,g,f_i)$ takes the form
	\begin{align*}
		\tilde{\chi}_t(\gamma_\mrc,g,f_i) = \cos(\sqrt{2f_i} gt)+ \frac{\frac{\gamma_\mrc}{g}}{\sqrt{32f_i}} \sin(\sqrt{2f_i} gt) 
	\end{align*}
	with $i=0,1,2$. This leads to the expression for the function $\phi$,
	\begin{align}
		&\phi(F,g,\gamma_\mrc,\tau) = \frac{1}{2(1+\frac{F^2}{g^2})} \left \{ \frac{8F^2}{g^2} \tilde{\chi}_\tau(\gamma_\mrc,g,f_0) \right . \nonumber\\
		&  \left . + \left(1+\sqrt{1+\frac{4F^2}{g^2}}\right) \tilde{\chi}_\tau(\gamma_\mrc,g,f_1) \right. \nonumber\\
		& \left. + \left(1-\sqrt{1+\frac{4F^2}{g^2}}\right) \tilde{\chi}_\tau(\gamma_\mrc,g,f_2) \right\}
		\label{eq:phifuncsmallg},
	\end{align}
	which contains only sinusoidal oscillatory terms. As a result, from Eq.~\eqref{eq:taugammaSmall1} it is clear that the exponential term $e^{-\gamma_\mrc \tau/4}$ alone determines the charging time given in Eq.~\eqref{eq:taugammaSmall2}. Note that for $\gamma_\mrc \ll g$ the charging time is independent of $F$ and $g$.
	
	Considering now the opposite limit of $\gamma_\mrc \gg \{F,g\}$, we can make the approximations:
	\begin{align}
		&2 \chi_t(\gamma_\mrc,g,f_j) \stackrel{\gamma_\mrc \gg \{ F,g \}}{\approx} e^{-\frac{4g^2f_j t}{\gamma_\mrc}}+e^{-\frac{\gamma_{\mrc}t}{2}}e^{\frac{4g^2f_jt}{\gamma_{\mrc}}} \nonumber \\
		& + \left( 1+\frac{16f_jg^2}{\gamma_{\mrc}^2}\right) \left ( e^{-\frac{4g^2f_j t}{\gamma_\mrc}}-e^{-\frac{\gamma_{\mrc}t}{2}}e^{\frac{4g^2f_jt}{\gamma_{\mrc}}} \right) \label{eq:chilargegammaC}.
	\end{align}
	Substituting this into Eq.~\eqref{eq:EnergyanalytFnTLSres} and keeping terms to order $\frac{1}{\gamma_\mrc}$, we can reduce the condition \eqref{eq:taudetexpcondn} to
	\begin{align}
		&\frac{1}{2\left(1+\frac{4F^2}{g^2}\right)}\left \vert \frac{8F^2}{g^2} e^{-\frac{2g^2}{\gamma_{\mrc}}\tau} + \left(1+\sqrt{1+\frac{4F^2}{g^2}}\right) \right . \nonumber\\
		&\left .e^{-\frac{4f_1 g^2}{\gamma_{\mrc}}\tau}+\left(1-\sqrt{1+\frac{4F^2}{g^2}}\right)e^{-\frac{4 f_2\, g^2}{\gamma_{\mrc}}\tau} \right \vert = e^{-n}
		\label{eq:taugammaLarge1}.
	\end{align}
	In contrast to the small dephasing limit, here we have three damping timescales and the charging time will be decided by the slowest scale among them. As we show in more detail in the SM, while for the weak driving $F/g \ll 1$ regime the slowest damping scale comes from the second term ($4 f_1/\gamma_\mrc$), in the strong driving $F/g\gg 1$ it is given by the first term ($2g^2/\gamma_\mrc$). Noting that in the limit of $F/g \ll 1$, we have $f_1 = 2\frac{F^4}{g^4} + O(F^6/g^6)$, and taking the inverse of the slowest damping timescale in the two regimes immediately leads to the results presented in Eqs.~\eqref{eq:tauweakdephasing} and \eqref{eq:taugammaLarge2}.
	\begin{figure}
    \centering
		\begin{overpic}[width=0.8\linewidth]{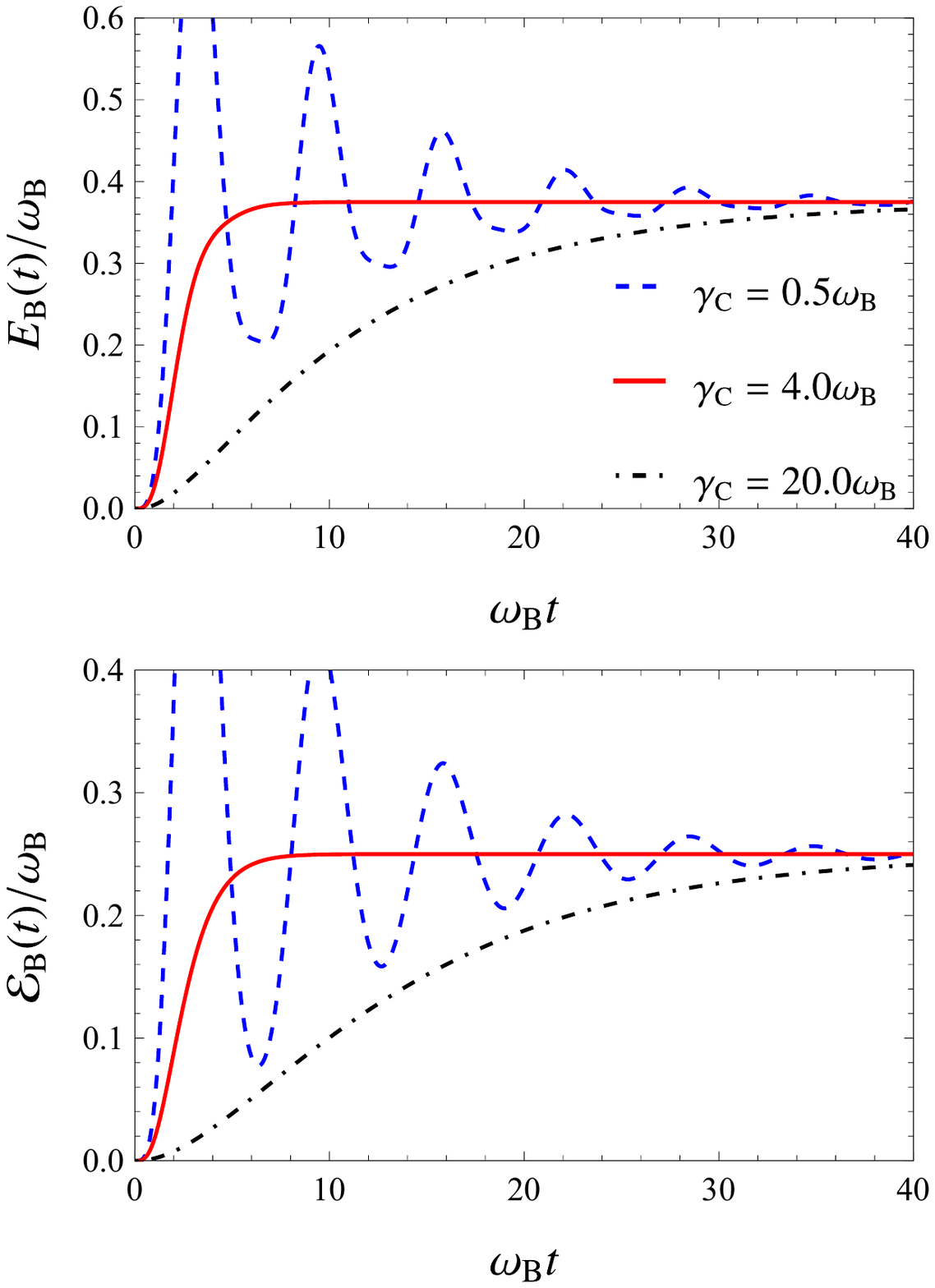}
		    \put(60,90){\textbf{(a)}}
            \put(60,40){\textbf{(b)}}
		\end{overpic}
		\caption{\textbf{Dynamics of energy and ergotropy for the two harmonic oscillator charger-battery system.} Time evolution of average energy $E_{\mr{B}}(t)$ [(a)] and ergotropy $\mathcal{E}_{\mr{B}}(t)$ [(b)] of the battery for the two-HO model for different values of the dephasing rate $\gamma_{\mr{C}}$. We show the resonant case, i.e. $\omega_\mrc = \omega_\mr{d} = \omega_\mrb$, for the optimal driving with $F=0.5 \omega_\mrb$, and $g = 1.0 \omega_\mrb$ ($F/g=0.5$).
		}
		\label{fig:figtwoHOmoderate}
	\end{figure}
	\begin{figure}
    \centering
		\begin{overpic}[width=0.8\linewidth]{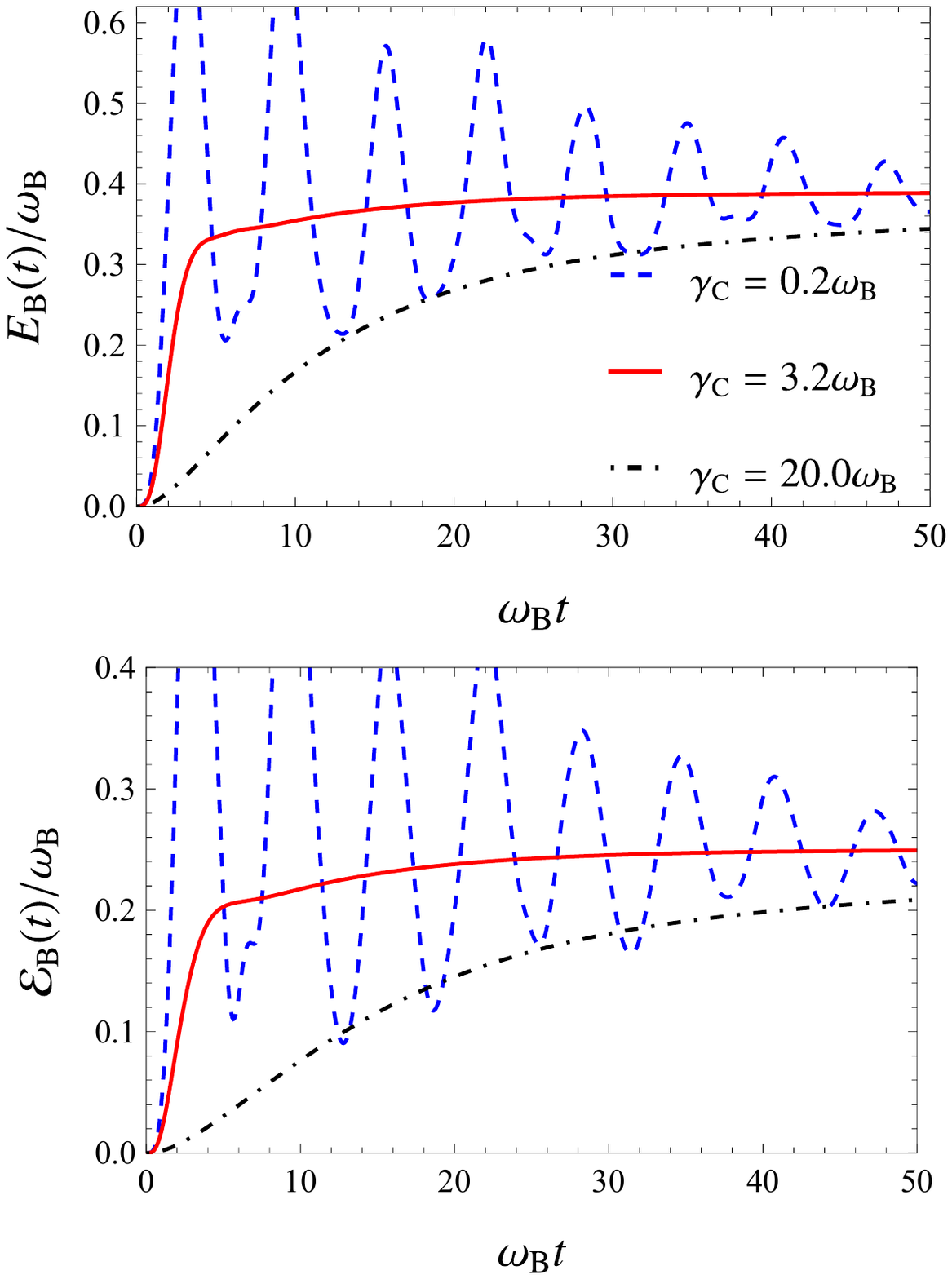}
		    \put(60,90){\textbf{(a)}}
            \put(60,40){\textbf{(b)}}
		\end{overpic}
		\caption{\textbf{Dynamics of energy and ergotropy for the hybrid TLS-harmonic oscillator system.} Time evolution of average energy $E_{\mr{B}}(t)$ [(a)] and ergotropy $\mathcal{E}_{\mr{B}}(t)$ [(b)] of the battery for the TLS (charger)-HO (battery) model for different values of the dephasing rate $\gamma_{\mr{C}}$. We show the resonant case, i.e. $\omega_\mrc = \omega_\mr{d} = \omega_\mrb$, for the optimal driving with $F=0.5 \omega_\mrb$, and $g = 1.0 \omega_\mrb$ ($F/g=0.5$).
		}
		\label{fig:figTLSHOmoderate}
	\end{figure}
    \modred{
	\noindent \subsubsection*{Results for two-HO and hybrid TLS-HO systems}
	\noindent To illustrate the generality of the central result presented in the paper, namely that of moderate dephasing leading to fast charging, we present the results with two additional setups. In the first case, the charger and battery are modeled as HOs and the Hamiltonian is given by
		\begin{align}
			\Hop =&\ \omega_\mrc \adop_{\mr{C}}\aop_{\mr{C}}+\omega_\mrb \adop_{\mr{B}}\aop_{\mr{B}} + g(\adop_{\mr{C}}\aop_\mrb + \adop_\mrb\aop_\mrc)\nonumber \\
			& + F(\aop_\mrc e^{i\omega_{\mr{d}} t} + \adop_\mrc e^{-i\omega_{\mr{d}} t}) \label{eq:bareHO},
		\end{align}
		with $\aop_\mrb$ and $\aop_\mrc$ denoting the annihilation operators for the battery and charger HO, respectively. The dephasing operator in Eq.~\eqref{eq:GKLSgen} is given by $\Lop_\mrc = \adop_\mrc \aop_\mrc$. While we present a detailed analysis of the dynamics of this setup in Section III of the SM, Fig.~\ref{fig:figtwoHOmoderate} illustrates how moderate dephasing of the charger HO leads to fast charging for the same parameters as the two-TLS case presented in Fig.~\ref{fig:figTLS2}.
		Finally, we consider a hybrid setting with a TLS charger and a HO battery with the Hamiltonian
		\begin{align}
			\Hop =&\ \omega_\mrc \hsp_{\mr{C}}\hsm_{\mr{C}}+\omega_\mrb \Bopd\Bop + g(\hsp_{\mr{C}}\Bop + \hsm_\mrc\Bopd)\nonumber \\
			& + F(\hsm_\mrc e^{i\omega_{\mr{d}} t} + \hsp_\mrc e^{-i\omega_{\mr{d}} t}) \label{eq:bareHTLSHO},
		\end{align}
		and the dephasing operator $\Lop_\mrc = \hsp_{\mr{C}}\hsm_{\mr{C}}$. Figure~\ref{fig:figTLSHOmoderate} illustrates our central result of dephasing enabled fast charging for this setup with a moderate driving $F/g = 0.5$. Similar results also hold for strong and weak driving (see Section IV of the SM).}

\begin{thebibliography}{10}
	\providecommand{\url}[1]{{#1}}
	\providecommand{\urlprefix}{URL }
	\providecommand{\doi}[1]{\url{https://doi.org/#1}}
	\bibcommenthead
	
	\bibitem{Quach2023}
	J.~Quach, G.~Cerullo, T.~Virgili, Quantum batteries: The future of energy
	storage?
	\newblock Joule \textbf{7}(10), 2195--2200 (2023).
	
	\bibitem{Bhattacharjee2021}
	S.~Bhattacharjee, A.~Dutta, Quantum thermal machines and batteries.
	\newblock Eur. Phys. J. B \textbf{94}(12), 239 (2021).
	
	\bibitem{myers2022quantum}
	N.M. Myers, O.~Abah, S.~Deffner, Quantum thermodynamic devices: From
	theoretical proposals to experimental reality.
	\newblock AVS Quantum Sci. \textbf{4}(2), 027101 (2022).
	\newblock
	
	\bibitem{Campaioli2018}
	F.~Campaioli, F.A. Pollock, S.~Vinjanampathy, \emph{Quantum Batteries}
	(Springer International Publishing, Cham, 2018), pp. 207--225.
	
	\bibitem{Andolina2018}
	G.M. Andolina, D.~Farina, A.~Mari, V.~Pellegrini, V.~Giovannetti, M.~Polini,
	Charger-mediated energy transfer in exactly solvable models for quantum
	batteries.
	\newblock Phys. Rev. B \textbf{98}, 205423 (2018).
	
	\bibitem{GarciaPintos2020}
	L.P. Garc\'{\i}a-Pintos, A.~Hamma, A.~del Campo, Fluctuations in extractable
	work bound the charging power of quantum batteries.
	\newblock Phys. Rev. Lett. \textbf{125}, 040601 (2020).
	\newblock
	
	\bibitem{Chen2022}
	P.~Chen, T.S. Yin, Z.Q. Jiang, G.R. Jin, Quantum enhancement of a single
	quantum battery by repeated interactions with large spins.
	\newblock Phys. Rev. E \textbf{106}, 054119 (2022).
	
	\bibitem{Dou2022}
	F.Q. Dou, Y.J. Wang, J.A. Sun, Highly efficient charging and discharging of
	three-level quantum batteries through shortcuts to adiabaticity.
	\newblock Front. Phys. \textbf{17}(3), 31503 (2021).
	
	\bibitem{Gyhm2023}
	J.Y. Gyhm, D.~Rosa, D.~\ifmmode~\check{S}\else \v{S}\fi{}afr\'anek, The minimal
	time it takes to charge a quantum system.
	\newblock arXiv:2308.16086  (2023).
	
	\bibitem{Campaioli2023}
	F.~Campaioli, S.~Gherardini, J.Q. Quach, M.~Polini, G.M. Andolina, Colloquium:
	Quantum batteries.
	\newblock arXiv:2308.02277  (2023).
	
	\bibitem{Hovhannisyan2013}
	K.V. Hovhannisyan, M.~Perarnau-Llobet, M.~Huber, A.~Ac\'{\i}n, Entanglement
	generation is not necessary for optimal work extraction.
	\newblock Phys. Rev. Lett. \textbf{111}, 240401 (2013).
	\newblock
	
	\bibitem{Gallacher2015}
	F.C. Binder, S.~Vinjanampathy, K.~Modi, J.~Goold, Quantacell: powerful charging
	of quantum batteries.
	\newblock New J. Phys. \textbf{17}(7), 075015 (2015).
	
	\bibitem{Campaioli2017}
	F.~Campaioli, F.A. Pollock, F.C. Binder, L.~C\'eleri, J.~Goold,
	S.~Vinjanampathy, K.~Modi, Enhancing the charging power of quantum batteries.
	\newblock Phys. Rev. Lett. \textbf{118}, 150601 (2017).
	\newblock
	
	\bibitem{Ferraro2018}
	D.~Ferraro, M.~Campisi, G.M. Andolina, V.~Pellegrini, M.~Polini, High-power
	collective charging of a solid-state quantum battery.
	\newblock Phys. Rev. Lett. \textbf{120}, 117702 (2018).
	\newblock
	
	\bibitem{Andolina2019}
	G.M. Andolina, M.~Keck, A.~Mari, M.~Campisi, V.~Giovannetti, M.~Polini,
	Extractable work, the role of correlations, and asymptotic freedom in quantum
	batteries.
	\newblock Phys. Rev. Lett. \textbf{122}, 047702 (2019).
	\newblock
	
	\bibitem{Ito2020}
	K.~Ito, G.~Watanabe, Collectively enhanced high-power and high-capacity
	charging of quantum batteries via quantum heat engines.
	\newblock arXiv:2008.07089  (2020).
	
	\bibitem{Rossini2020}
	D.~Rossini, G.M. Andolina, D.~Rosa, M.~Carrega, M.~Polini, Quantum advantage in
	the charging process of sachdev-ye-kitaev batteries.
	\newblock Phys. Rev. Lett. \textbf{125}, 236402 (2020).
	\newblock
	
	\bibitem{Julia-Farre2020}
	S.~Juli\`a-Farr\'e, T.~Salamon, A.~Riera, M.N. Bera, M.~Lewenstein, Bounds on
	the capacity and power of quantum batteries.
	\newblock Phys. Rev. Res. \textbf{2}, 023113 (2020).
	\newblock
	
	\bibitem{Watanabe2020}
	G.~Watanabe, B.P. Venkatesh, P.~Talkner, M.J. Hwang, A.~del Campo, Quantum
	statistical enhancement of the collective performance of multiple bosonic
	engines.
	\newblock Phys. Rev. Lett. \textbf{124}, 210603 (2020).
	\newblock
	
	\bibitem{Gyhm2022}
	J.Y. Gyhm, D.~\ifmmode~\check{S}\else \v{S}\fi{}afr\'anek, D.~Rosa, Quantum
	charging advantage cannot be extensive without global operations.
	\newblock Phys. Rev. Lett. \textbf{128}, 140501 (2022).
	\newblock
	
	\bibitem{Shaghaghi2022}
	V.~Shaghaghi, V.~Singh, G.~Benenti, D.~Rosa, {Micromasers as quantum
		batteries}.
	\newblock Quantum Sci. Technol. \textbf{7}, 04LT01 (2022).
	\newblock
	
	\bibitem{Yan2023}
	J.S. Yan, J.~Jing, Charging by quantum measurement.
	\newblock Phys. Rev. Appl. \textbf{19}, 064069 (2023).
	\newblock
	
	\bibitem{Yang2023a}
	D.L. Yang, F.M. Yang, F.Q. Dou, Three-level dicke quantum battery.
	\newblock arXiv:2308.01188  (2023).
	
	\bibitem{Gyhm2024}
	J.Y. Gyhm, U.R. Fischer, {Beneficial and detrimental entanglement for quantum
		battery charging}.
	\newblock AVS Quantum Sci. \textbf{6}(1), 012001 (2024).
	
	\bibitem{Liu2019}
	J.~Liu, D.~Segal, G.~Hanna, Loss-free excitonic quantum battery.
	\newblock J. Phys. Chem. C \textbf{123}(30), 18303--18314 (2019).
	
	\bibitem{Pirmoradian2019}
	F.~Pirmoradian, K.~M\o{}lmer, Aging of a quantum battery.
	\newblock Phys. Rev. A \textbf{100}, 043833 (2019).
	
	\bibitem{Barra2019}
	F.~Barra, Dissipative charging of a quantum battery.
	\newblock Phys. Rev. Lett. \textbf{122}, 210601 (2019).
	\newblock
	
	\bibitem{Farina2019}
	D.~Farina, G.M. Andolina, A.~Mari, M.~Polini, V.~Giovannetti, Charger-mediated
	energy transfer for quantum batteries: An open-system approach.
	\newblock Phys. Rev. B \textbf{99}, 035421 (2019).
	
	\bibitem{Quach2020}
	J.Q. Quach, W.J. Munro, Using dark states to charge and stabilize open quantum
	batteries.
	\newblock Phys. Rev. Appl. \textbf{14}, 024092 (2020).
	\newblock
	
	\bibitem{Tabesh2020}
	F.T. Tabesh, F.H. Kamin, S.~Salimi, Environment-mediated charging process of
	quantum batteries.
	\newblock Phys. Rev. A \textbf{102}, 052223 (2020).
	
	\bibitem{Kamin2020}
	F.H. Kamin, F.T. Tabesh, S.~Salimi, F.~Kheirandish, A.C. Santos, Non-markovian
	effects on charging and self-discharging process of quantum batteries.
	\newblock New J. Phys. \textbf{22}, 083007 (2020).
	\newblock
	
	\bibitem{Mitchison2021chargingquantum}
	M.T. Mitchison, J.~Goold, J.~Prior, Charging a quantum battery with linear
	feedback control.
	\newblock {Quantum} \textbf{5}, 500 (2021).
	
	\bibitem{Xu2021}
	K.~Xu, H.J. Zhu, G.F. Zhang, W.M. Liu, Enhancing the performance of an open
	quantum battery via environment engineering.
	\newblock Phys. Rev. E \textbf{104}, 064143 (2021).
	
	\bibitem{Santos2021}
	A.C. Santos, Quantum advantage of two-level batteries in the self-discharging
	process.
	\newblock Phys. Rev. E \textbf{103}, 042118 (2021).
	
	\bibitem{Ghosh2021}
	S.~Ghosh, T.~Chanda, S.~Mal, A.~Sen(De), Fast charging of a quantum battery
	assisted by noise.
	\newblock Phys. Rev. A \textbf{104}, 032207 (2021).
	
	\bibitem{Quach2022}
	J.Q. Quach, K.E. McGhee, L.~Ganzer, D.M. Rouse, B.W. Lovett, E.M. Gauger,
	J.~Keeling, G.~Cerullo, D.G. Lidzey, T.~Virgili, Superabsorption in an
	organic microcavity: Toward a quantum battery.
	\newblock Sci. Adv. \textbf{8}(2), eabk3160 (2022).
	\newblock
	
	\bibitem{Mayo2022}
	F.~Mayo, A.J. Roncaglia, Collective effects and quantum coherence in
	dissipative charging of quantum batteries.
	\newblock Phys. Rev. A \textbf{105}, 062203 (2022).
	
	\bibitem{Rodriguez2022}
	R.R. Rodriguez, B.~Ahmadi, G.~Suarez, P.~Mazurek, S.~Barzanjeh, P.~Horodecki,
	Optimal quantum control of charging quantum batteries.
	\newblock arXiv:2207.00094  (2023).
	
	\bibitem{Shaghaghi2023}
	V.~Shaghaghi, V.~Singh, M.~Carrega, D.~Rosa, G.~Benenti, Lossy micromaser
	battery: Almost pure states in the jaynes–cummings regime.
	\newblock Entropy \textbf{25}, 430 (2023).
	
	\bibitem{Yang2023b}
	F.M. Yang, F.Q. Dou, Resonator-qutrits quantum battery.
	\newblock arXiv:2312.11006  (2023).
	
	\bibitem{Rodriguez2023}
	R.R. Rodr\'{\i}guez, B.~Ahmadi, P.~Mazurek, S.~Barzanjeh, R.~Alicki,
	P.~Horodecki, Catalysis in charging quantum batteries.
	\newblock Phys. Rev. A \textbf{107}, 042419 (2023).
	
	\bibitem{Dou2023}
	F.Q. Dou, F.M. Yang, Superconducting transmon qubit-resonator quantum battery.
	\newblock Phys. Rev. A \textbf{107}, 023725 (2023).
	
	\bibitem{Ahmadi2024}
	B.~Ahmadi, P.~Mazurek, P.~Horodecki, S.~Barzanjeh, Nonreciprocal quantum
	batteries.
	\newblock arXiv:2401.05090  (2024).
	
	\bibitem{gangwar2024coherently}
	K.~Gangwar, A.~Pathak, Coherently driven quantum harmonic oscillator battery.
	\newblock arXiv:2401.07238  (2024).
	
	\bibitem{Gherardini2020}
	S.~Gherardini, F.~Campaioli, F.~Caruso, F.C. Binder, Stabilizing open quantum
	batteries by sequential measurements.
	\newblock Phys. Rev. Research \textbf{2}, 013095 (2020).
	\newblock
	
	\bibitem{QSLRef_Deffner_2017}
	S.~Deffner, S.~Campbell, Quantum speed limits: from heisenberg’s uncertainty
	principle to optimal quantum control.
	\newblock J. Phys. A \textbf{50}(45), 453001 (2017).
	
	\bibitem{QSLRef_Funo}
	K.~Funo, N.~Shiraishi, K.~Saito, Speed limit for open quantum systems.
	\newblock New J. Phys. \textbf{21}(1), 013006 (2019).
	
	\bibitem{QSLRef_Pati2}
	C.~Mukhopadhyay, A.~Misra, S.~Bhattacharya, A.K. Pati, Quantum speed limit
	constraints on a nanoscale autonomous refrigerator.
	\newblock Phys. Rev. E \textbf{97}, 062116 (2018).
	
	\bibitem{nielsen2010quantum}
	M.A. Nielsen, I.L. Chuang, \emph{Quantum computation and quantum information}
	(Cambridge University Press, Cambridge, 2010)
	
	\bibitem{Skinner86}
	J.L. Skinner, D.~Hsu, Pure dephasing of a two-level system.
	\newblock J. Phys. Chem. \textbf{90}(21), 4931--4938 (1986).
	
	\bibitem{Lidar2001}
	D.A. Lidar, Z.~Bihary, K.~Whaley, From completely positive maps to the quantum
	markovian semigroup master equation.
	\newblock Chem. Phys. \textbf{268}(1), 35--53 (2001).
	\newblock
	
	\bibitem{Albash2015}
	T.~Albash, D.A. Lidar, Decoherence in adiabatic quantum computation.
	\newblock Phys. Rev. A \textbf{91}, 062320 (2015).
	
	\bibitem{Steck24}
	D.A.~Steck, \emph{Quantum and Atom Optics} (available online at
	\href{http://steck.us/teaching}{http://steck.us/teaching} (revision 0.16.1,
	16 June 2024))
	
	\bibitem{wiseman2009quantum}
	H.M. Wiseman, G.J. Milburn, \emph{Quantum measurement and control} (Cambridge
	University Press, Cambridge, 2009)
	
	\bibitem{Dorner_2012}
	U.~Dorner, Quantum frequency estimation with trapped ions and atoms.
	\newblock New J. Phys. \textbf{14}(4), 043011 (2012).
	
	\bibitem{ShastriFuture}
	R.~Shastri, C.~Jiang, G.H.~Xu, B.~Prasanna Venkatesh, G.~Watanabe (unpublished)
	
	\bibitem{Francica2020}
	G.~Francica, F.C. Binder, G.~Guarnieri, M.T. Mitchison, J.~Goold, F.~Plastina,
	Quantum coherence and ergotropy.
	\newblock Phys. Rev. Lett. \textbf{125}, 180603 (2020).
	\newblock
	
	\bibitem{Qutip}
	J.~Johansson, P.~Nation, F.~Nori, Qutip 2: A python framework for the dynamics
	of open quantum systems.
	\newblock Comput. Phys. Commun. \textbf{184}(4), 1234--1240 (2013).
	\newblock
	
	\bibitem{Hu2022}
	C.K. Hu, J.~Qiu, P.J.P. Souza, J.~Yuan, Y.~Zhou, L.~Zhang, J.~Chu, X.~Pan,
	L.~Hu, J.~Li, Y.~Xu, Y.~Zhong, S.~Liu, F.~Yan, D.~Tan, R.~Bachelard, C.J.
	Villas-Boas, A.C. Santos, D.~Yu, Optimal charging of a superconducting
	quantum battery.
	\newblock Quantum Sci. Technol. \textbf{7}(4), 045018 (2022).
	
	\bibitem{Joshi2022}
	J.~Joshi, T.S. Mahesh, Experimental investigation of a quantum battery using
	star-topology nmr spin systems.
	\newblock Phys. Rev. A \textbf{106}, 042601 (2022).
	
	\bibitem{Berger2015measurement}
	S.~Berger, M.~Pechal, P.~Kurpiers, A.A. Abdumalikov, C.~Eichler, J.A. Mlynek,
	A.~Shnirman, Y.~Gefen, A.~Wallraff, S.~Filipp, Measurement of geometric
	dephasing using a superconducting qubit.
	\newblock Nat. Commun. \textbf{6}(1), 8757 (2015).
	
	\bibitem{breuer_theory_2007}
	H.P. Breuer, F.~Petruccione, \emph{The {Theory} of {Open} {Quantum} {Systems}}
	(Oxford University Press, 2007).
	\newblock
	
	\bibitem{Mathematica}
	{Wolfram Research Inc.}
	\newblock {Mathematica, {V}ersion 12.0}.
	\newblock {Champaign, IL, 2020}
	
\end{thebibliography}
\noindent \section*{Data Availability}
\noindent All data generated during this study can be reproduced using the described methodology

\noindent \section*{Code Availability}
\noindent Numerical codes used to generate some of the results can be provided upon reasonable request.
\section*{Acknowledgements}
\noindent G.W. was supported by the National Natural Science Foundation of China (Grants No. 12375039 and No. 11975199). B.P.V. acknowledges the
MATRICS Grant No.~MTR/2023/000900 from Science and Engineering Research Board/Anusandhan National Research Foundation, Government of India.
\section*{Author Contributions}
G.W. and B.P.V. conceived the project. R.S., B.P.V., and G.W. conducted theoretical and numerical studies. All the authors participated in the discussions. B.P.V., G.W., and R.S. wrote the manuscript with input from the other authors. B.P.V. and G.W. led the project.
\section*{Competing Interests}
The authors declare no competing interests
\section*{Additional Information}
\textbf{Supplementary information} In the supplementary material we provide the details regarding the derivation of the GKLS master equation, analytical calculations and dynamics with detuning for the TLS battery-charger setup, and additional details of the results with HO and hybrid TLS-HO setups. It also includes Refs.~\cite{Steck24,wiseman2009quantum,Dorner_2012,Farina2019,ShastriFuture,Francica2020,gangwar2024coherently,Qutip}.

\section*{Figure Legends}
\begin{itemize}
\item \textbf{Schematic of the Charger-Battery setup--} Schematic of the setup of a quantum battery B coupled to a quantum charger system C with a coupling constant $g$. The charger is driven at a rate $F$ and additionally subject to dephasing at the rate $\gamma_\mrc$.
\item \textbf{Dynamics of Energy and Ergotropy of the battery for the two-TLS system --} Time evolution of average energy $E_{\mr{B}}(t)$ [(a)] and ergotropy $\mathcal{E}_{\mr{B}}(t)$ [(b)] of the battery for the two-TLS model for different values of the dephasing rate $\gamma_{\mr{C}}$. Here, as an example, we show the resonant case, i.e. $\omega_\mrc = \omega_\mr{d} = \omega_\mrb$, for the optimal driving with $F=0.5 \omega_\mrb$, and $g = 1.0 \omega_\mrb$ ($F/g=0.5$).
\item \textbf{Charging time as a function of dephasing--} Charging time $\tau$ of the battery for the two-TLS model as a function of the charger dephasing rate $\gamma_\mrc$ for different regimes of the driving strength: (a) weak driving $F=0.1 \omega_\mrb$, $g = \omega_\mrb$ ($F/g=0.1$), (b) strong driving $F=10.0 \omega_\mrb$, $g = \omega_\mrb$ ($F/g=10.0$), and (c) optimal driving $F=0.5 \omega_\mrb$, $g = \omega_\mrb$ ($F/g=0.5$). Here, we show the resonant case, i.e., $\omega_\mrc = \omega_\mr{d} = \omega_\mrb$. Gray vertical lines indicate the value, $\gamma_\mrc^\star$, of the optimal dephasing to get the fastest charging time.
\item \textbf{Energy of the battery coupled to a detuned charger--}(a) Transient dynamics of the average energy $E_{\mrb}(t)$ of the battery in the weak driving regime $F/g=0.1$ for the detuned charger-battery case with $\Delta_{\mr{CB}}=0.03 \omega_{\mrb}$. (b) Maximum value of the oscillating energy, $E_\mrb^{\mr{max}}$, in the closed case (blue dashed line) is compared to the case with dephasing (red line) as function of the charger-battery detuning. Other parameter values are $g=\omega_{\mrb}$ and $\omega_{\mr{d}}=\omega_\mrb$.
\item \textbf{Dynamics of Energy and Ergotropy for the two harmonic oscillator charger-battery system--}Time evolution of average energy $E_{\mr{B}}(t)$ [(a)] and ergotropy $\mathcal{E}_{\mr{B}}(t)$ [(b)] of the battery for the two-HO model for different values of the dephasing rate $\gamma_{\mr{C}}$. We show the resonant case, i.e. $\omega_\mrc = \omega_\mr{d} = \omega_\mrb$, for the optimal driving with $F=0.5 \omega_\mrb$, and $g = 1.0 \omega_\mrb$ ($F/g=0.5$).
\item \textbf{Dynamics of energy and ergotropy for the hybrid TLS-harmonic oscillator system.} Time evolution of average energy $E_{\mr{B}}(t)$ [(a)] and ergotropy $\mathcal{E}_{\mr{B}}(t)$ [(b)] of the battery for the TLS (charger)-HO (battery) model for different values of the dephasing rate $\gamma_{\mr{C}}$. We show the resonant case, i.e. $\omega_\mrc = \omega_\mr{d} = \omega_\mrb$, for the optimal driving with $F=0.5 \omega_\mrb$, and $g = 1.0 \omega_\mrb$ ($F/g=0.5$).
\end{itemize}
\end{document}


	
\title{--- Supplemental Material --- \texorpdfstring{\\Dephasing Enabled Fast Charging of Quantum Batteries}{maintitle}}
\author{Rahul Shastri}
\affiliation{Indian Institute of Technology Gandhinagar, Palaj, Gujarat 382055, India}
\author{Chao Jiang}
\affiliation{School of Physics and Zhejiang Institute of Modern Physics, Zhejiang University, Hangzhou, Zhejiang 310027, China}
\affiliation{Graduate School of China Academy of Engineering Physics, Beijing 100193, China}
\author{Guo-Hua Xu}
\affiliation{School of Physics and Zhejiang Institute of Modern Physics, Zhejiang University, Hangzhou, Zhejiang 310027, China}
\author{B. Prasanna Venkatesh}
\email{prasanna.b@iitgn.ac.in}
\affiliation{Indian Institute of Technology Gandhinagar, Palaj, Gujarat 382055, India}
\author{Gentaro Watanabe}
\email{gentaro@zju.edu.cn}
\affiliation{School of Physics and Zhejiang Institute of Modern Physics, Zhejiang University, Hangzhou, Zhejiang 310027, China}
\affiliation{Zhejiang Province Key Laboratory of Quantum Technology and Device, Zhejiang University, Hangzhou, Zhejiang 310027, China}
	
\maketitle

\modred{\section{\label{sec:MasterEqn} Master Equation Derivation}

In this section, we provide two approaches to derive the GKLS master equation (1) of the main paper for the dephased charger-battery system. In both approaches, we start with a charger-batter system with a total Hamiltonian $\Hop(t)= \Hop_{\mr{C}} +\Hop_{\mr{d}}(t) + \Hop_{\mr{B}} + \Hop_{\mr{CB}}$ describing the driven charger coupled to a battery. This Hamiltonian is time dependent due to the driving but as will be evident from the specific set-ups we have considered in the main paper, it is possible to move to a transformed frame via a unitary operator $\Uop_{\mr{d}}(t)$ that leads to a time-independent Hamiltonian $\olsi{\Hop} = \Uop_{\mr{d}}(t) \Hop(t) \Uop_{\mr{d}}^{\dagger}(t)$ describing the system. For specific forms of the unitary $\Uop_{\mr{d}}(t)$ for the two-TLS and two-HO set-ups please see Eqs.~\eqref{eq:UdTLS} and \eqref{eq:UdHO} respectively. We will work with the time-independent Hamiltonian $\olsi{\Hop}$ in the following derivations. In the end, we can undo the unitary and come back to the `lab' frame in which Eq.~(1) of the main paper is given.

\subsection*{Dephasing Master Equation from Continuous Quantum Measurement}
The idea in this approach is to perform a continuous weak measurement of the observable $\Lop_\mrc$ that will eventually correspond to the 
dephasing operator \cite{Steck24,wiseman2009quantum}. To this end let us consider the Gaussian (and hermitian) POVM operators with outcomes $\alpha$ defined as
\begin{align}
    \hat{\Omega}(\alpha,\Delta t) = \left( \frac{2\gamma_\mrc \Delta t}{\pi}\right)^{1/4} \sum_{m} e^{-\gamma_\mrc \Delta t (\lambda_m-\alpha)^2} \hat{\Pi}_{m}, \label{eq:POVM}
\end{align}
where $\hat{\Pi}_{m}$ are eigenprojectors of the operator $\Lop_{\mrc}$ with eigenvalues $\lambda_m$, i.e., $\Lop_{\mrc} \hat{\Pi}_{m} = \lambda_m \hat{\Pi}_{m}$ and $\Delta t>0$ represents a time step that will eventually be made infinitesimal. Note that we have assumed that the operator $\Lop_\mrc$ has a discrete spectrum for simplicity. In order to derive the equation describing the time evolution of the system under continuous measurement, we Trotterize the time evolution with time steps $\Delta t$ and intersperse the unitary evolution of the system by the Hamiltonian $\olsi{\Hop}$ repeatedly with measurements realised via the POVM operators given in Eq.~\eqref{eq:POVM}, and eventually take the limit $\Delta t \rightarrow 0$. The probability of outcome $\alpha$ for the POVM given the system state $\hrho(t)$ is
\begin{align*}
    P(\alpha) &= \mr{Tr} [\hat{\Omega}(\alpha,\Delta t) \hrho \hat{\Omega}(\alpha,\Delta t)]\\
    &= \sqrt{\frac{2\gamma_\mrc \Delta t}{\pi}} \sum_{m} e^{-2\gamma_\mrc \Delta t (\lambda_m-\alpha)^2} \mr{Tr}\left [ \hat{\Pi}_m \hrho(t)\right].
\end{align*}
In the limit of $\Delta t \rightarrow 0$, the Gaussian factor in the above sum becomes very broad and one can take it out of the sum with the replacement $\lambda_m \rightarrow \avg{\Lop_\mrc}$ \cite{Steck24}, leading the outcome probability to reduce to
\begin{align*}
    P(\alpha) \approx \sqrt{\frac{2 \gamma_\mrc \Delta t}{\pi}} e^{-2\gamma_\mrc\Delta t\left(\avg{\Lop_\mrc}-\alpha\right)^2}.
\end{align*}
In this limit, we can write the outcome variable as a Gaussian stochastic variable of the form:
\begin{align}
    \alpha = \avg{\Lop_\mrc} + \frac{\Delta W}{\sqrt{4\gamma_\mrc}\Delta t} \label{eq:alphagaussvar},
\end{align}
with $\Delta W$ denoting the Wiener increment. The time evolution of the system over a single time step, when starting from an initial pure state $\ket{\Psi(t)}$, due to the measurement (with outcome $\alpha$) and unitary evolution can be written as
\begin{align}
\ket{\Psi(t+\Delta t)} = \frac{e^{-i\olsi{\Hop} \Delta t} \hat{\Omega}(\alpha,\Delta t) \ket{\Psi(t)}}{\sqrt{\sandwich{\Psi(t)}{\hat{\Omega}(\alpha,\Delta t)^2}{\Psi(t)}}}. 
\end{align}
Note that even if the measurement operator and the Hamiltonian do not necessarily commute, i.e., $[\Hop,\hat{\Omega}(\alpha,\Delta t)] \neq 0$, the above sequential or Trotterized evolution becomes accurate and valid as $\Delta t \rightarrow 0$. Ignoring the denominator and pre-factors that give the normalization, the non-normalized wave function $\ket{\tilde{\Psi}(t+\Delta t)}$ after a single time step using the stochastic form of $\alpha$ from Eq.~\eqref{eq:alphagaussvar} becomes
\begin{align*}
\ket{\tilde{\Psi}(t+\Delta t)} =&\ e^{-i\olsi{\Hop} \Delta t} \exp \left [-\gamma_\mrc \Delta t \Lop_\mrc  \right .\\
&\left . + \Lop_\mrc\left(2\gamma_\mrc \avg{\Lop_\mrc} \Delta t + \sqrt{\gamma_\mrc} \Delta W \right) \right] \ket{\Psi(t)}.
\end{align*}
Expanding the exponential to $O(\Delta)$ (keeping track of $\Delta W \sim O(\sqrt{\Delta t})$) and taking the infinitesimal limit with $\Delta t\rightarrow dt$ and $\Delta W \rightarrow dW$ (with $dW$ denoting the Wiener increment) results in
\begin{align*}
    \ket{\tilde{\Psi}(t+\Delta t)} =&\ \bigg[ I - i \olsi{\Hop} dt - \left(\frac{\gamma_\mrc}{2}\Lop_\mrc^2 - 2 \gamma_\mrc \Lop_\mrc \avg{\Lop_\mrc}\right)dt  \\
    & + \sqrt{\gamma_\mrc} \Lop_\mrc dW \bigg] \ket{\Psi(t)}.
\end{align*}
Normalizing $\ket{\tilde{\Psi} (t+\Delta t)}$ leads to the non-linear stochastic Sch\"{o}dinger equation
\begin{align}
    d \ket{\Psi} =&\ - i \olsi{\Hop} \ket{\Psi} dt - \frac{\gamma_\mrc}{2} \left(\Lop_\mrc-\avg{\Lop_\mrc} \right)^2 \ket{\Psi} dt \nonumber \\
    & \sqrt{\gamma_\mrc} \left (\Lop_\mrc-\avg{\Lop_\mrc} \right) \ket{\Psi} dW \label{eq:contmeasSSE}.
\end{align}
Taking $\hrho(t) = \ketbra{\Psi}{\Psi}$ and using $d \hrho = (d\ket{\Psi})\bra{\Psi} + \ket{\Psi}(d\bra{\Psi}) + (d \ket{\Psi})(d \bra{\Psi})$, we can write the stochastic master equation for the density matrix as
\begin{align}
    d \hrho &= -i \left[\olsi{\Hop},\hrho\right] dt + \gamma_\mrc \left(\Lop_\mrc \hrho(t) \Lop_\mrc - \frac{\{ \Lop_\mrc^2,\hrho(t)\}}{2} \right) dt \nonumber \\
    &+\sqrt{\gamma_\mrc} \left(\Lop_\mrc \hrho + \hrho \Lop_\mrc - 2 \avg{\Lop_\mrc} \hrho\right) dW \label{eq:contmeasSME}.
\end{align}
Averaging over the noise terms or many quantum trajectories corresponding to the stochastic master equation (noticing $\avg{dW} = 0$), and transforming back to the lab frame, we immediately obtain the GKLS master equation~(1) from the main paper. The final step of undoing the unitary transformation is enabled by the fact that in all the examples we consider $[\Uop_{\mr{d}}(t),\Lop_\mrc] = 0$ (see Eqs.~\eqref{eq:UdTLS} and \eqref{eq:UdHO}). 

\subsection*{Dephasing Master Equation from Classical Noise Model}
The second approach we present to derive the master equation involves adding the relevant dephasing operator $\Lop_\mrc$ (which is hermitian) multiplied by a classical noise term to the total Hamiltonian \cite{Dorner_2012}. For the TLS charger with $\Lop_\mrc \propto \hs^+_{\mr{C}} \hs^-_{\mr{C}}$, if the TLS is a spin qubit for example, one can motivate such a term by adding magnetic field noise. Similarly for a HO charger with $\Lop_\mrc \propto \adop_{\mr{C}} \aop_{\mr{C}}$, adding frequency noise to the oscillator can generate the same term. In any case, the Hamiltonian with the noise term becomes
\begin{align}
	\Hop_{\mr{n}} = \olsi{\Hop} +\sqrt{\gamma_\mrc} \xi(t)\Lop_{\mrc}, \label{eq:noisyH}
\end{align}
with $\xi(t) = \frac{dW}{dt}$ given by the standard white noise with properties (with the $\langle \langle {\cdot} \rangle \rangle $ symbol denoting noise averaging):
\begin{align*}
	\langle \langle\xi(t)\rangle \rangle &= 0,\\
	\langle \langle\xi(t) \xi(t^\prime)\rangle \rangle &= \delta(t-t^\prime).
\end{align*}
The time evolution of the charger-battery system under the above noisy Hamiltonian leads to a Stratonovich stochastic differential equation for the quantum state of the form:
\begin{align}
	d \ket{\Psi} = -i \olsi{\Hop} \ket{\Psi} dt - i \sqrt{\gamma_\mrc} \Lop_{\mrc} \ket{\Psi} dW. \label{eq:StratSDEnoise}
\end{align}
Converting this to the It\^o form, we arrive at a linear stochastic Schr\"{o}dinger equation and the associated stochastic master equation of the form:
\begin{align}
	d \ket{\Psi} &= \left(-i \olsi{\Hop} -\frac{\gamma_\mrc}{2} \Lop_{\mrc}\right)\ket{\Psi}dt - i \sqrt{\gamma_\mrc} \Lop_{\mrc} \ket{\Psi} dW,
	\label{eq:ItoSDEnoise}\\
	d \hrho =& -i \left [\olsi{\Hop},\hrho \right] dt + \frac{\gamma_\mrc}{2} \left(2\Lop_{\mrc}\hrho \Lop_{\mrc} - \Lop_{\mrc} \hrho -\hrho \Lop_{\mrc} \right) dt \nonumber \\ &-i \sqrt{\gamma_\mrc} \left( \Lop_{\mrc} \hrho - \hrho \Lop_{\mrc} \right) dW,
	\label{eq:ItoSMEnoise}
\end{align}
respectively. Again, averaging over the noise terms in Eq.~\eqref{eq:ItoSMEnoise} and rotating back to the lab frame immediately leads to the master equation (1) in the main paper. 
}

\section{\label{sec:A} Two-TLS Model}
In this section, we provide additional details pertaining to the dynamics of the two TLS charger-battery system discussed in the main paper. We begin by first noting that the analytical considerations of the two-TLS model are simplified if we rewrite the equations of motion for the moments given in Appendix A of the main paper in Eqs.~(11)--(13) into a time-independent form. This can be done by the transformations $\hs^-_{\mr{B}} \rightarrow \hs^-_{\mr{B}} e^{i(\omega_\mrb-\omega_\mr{d})t}$ and $\hs^-_{\mr{C}} \rightarrow \hs^-_{\mr{C}} e^{i(\omega_\mrc-\omega_\mr{d})t}$ at the level of the operators resulting in
\begin{align}
    \frac{d}{dt}\avg{\hs^z_{\mr{B}}} =& -4g\mathfrak{Im}\left[\avg{\hs^+_{\mr{C}}\hs^-_\mrb}\right], \nonumber\\
    \frac{d}{dt}\avg{\hs^z_{\mr{C}}} =& -4F \mathfrak{Im}\left[\avg{\hs^-_{\mr{C}}}\right] + 4g \mathfrak{Im} \left[\avg{\hs^+_{\mr{C}}\hs^-_{\mr{B}}}\right] \nonumber, \\
    \frac{d}{dt} \avg{\hs^+_{\mr{C}}\hs^-_{\mr{B}}} =& -iF\avg{\hs^z_{\mr{C}}\hs^-_{\mr{B}}} -ig\left(\avg{\hs^z_{\mr{C}}} - \avg{\hs^z_{\mr{B}}} \right)/2 \nonumber\\
    & -\left[\frac{\gamma_{\mr{C}}}{2} - i\left(\Delta_\mr{Cd}-\Delta_\mr{Bd}\right)\right]\avg{\hs^+_{\mr{C}}\hs^-_{\mr{B}}}, \nonumber \\ 
     \frac{d}{dt}\avg{\hs^z_{\mr{C}}\hs^-_{\mr{B}}} =& -2iF\left(\avg{\hs^+_{\mr{C}}\hs^-_{\mr{B}}} -  \avg{\hs^-_{\mr{C}}\hs^-_{\mr{B}}}\right) + ig\avg{\hs^-_{\mr{C}}} \nonumber\\
    &- i\Delta_\mr{Bd}\avg{\hs^z_\mrc \hs^-_\mrb}, \nonumber\\
    \frac{d}{dt}\avg{\hs^-_{\mr{C}}\hs^-_{\mr{B}}} =& iF\avg{\hs^z_{\mr{C}}\hs^-_{\mr{B}}} \nonumber\\
    & - \left[\frac{\gamma_{\mr{C}}}{2} + i\left(\Delta_\mr{Cd}+\Delta_\mr{Bd}\right)\right]\avg{\hs^-_{\mr{C}}\hs^-_{\mr{B}}}, \nonumber \\
    \frac{d}{dt}\avg{\hs^-_{\mr{C}}} =& iF\avg{\hs^z_{\mr{C}}} + ig\avg{\hs^z_{\mr{C}}\hs^-_{\mr{B}}}  - \left(\frac{\gamma_{\mr{C}}}{2} + i\Delta_\mr{Cd}\right)\avg{\hs^-_{\mr{C}}},
    \label{eq:TwoTLSEOMv2set1}
\end{align}
and 
\begin{align}
    \frac{d}{dt}\avg{\hs^-_{\mr{B}}} =&\ ig\avg{\hs^-_{\mr{C}}\hs^z_{\mr{B}}} -i\Delta_\mr{Bd}\avg{\hs^-_\mrb}, \nonumber\\
    \frac{d}{dt}\avg{\hs^-_{\mr{C}}\hs^z_{\mr{B}}} =&\ iF\avg{\hs^z_{\mr{C}}\hs^z_{\mr{B}}}- \left(\frac{\gamma_{\mr{C}}}{2} + i\Delta_\mr{Cd}\right)\avg{\hs^-_{\mr{C}}\hs^z_{\mr{B}}}, \nonumber\\
    & + ig\avg{\hs^-_{\mr{B}}} \nonumber\\
    \frac{d}{dt}\avg{\hs^z_{\mr{C}}\hs^z_{\mr{B}}} =& -4F\mathfrak{Im}\left[\avg{\hs^-_{\mr{C}}\hs^z_{\mr{B}}}\right]. \label{eq:TwoTLSEOMv2set2}
\end{align}
\begin{figure}
\subfloat{\begin{overpic}[width=0.85 \linewidth]{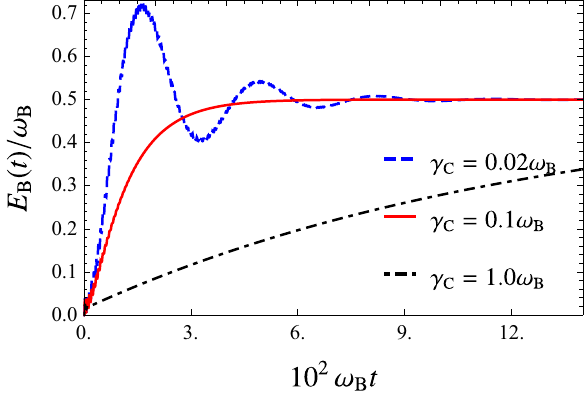}
	\put(80,58){\textbf{(a)}}
	\end{overpic}
	}\\
\subfloat{\begin{overpic}[width=0.85 \linewidth]{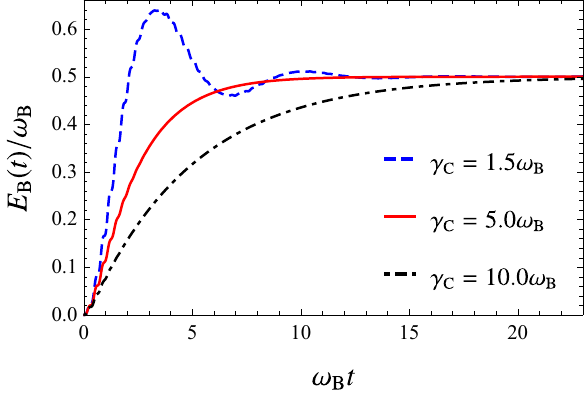}
	\put(80,58){\textbf{(b)}}
	\end{overpic}
	}
 \caption{(Color Online) Time evolution of the energy $E_{\mr{B}}(t)$ of the battery for two-TLS model for different values of the dephasing rate $\gamma_{\mr{C}}$. Panel (a) is for the weak driving case with parameters $F=0.1 \omega_\mrb$ and $g = \omega_\mrb$, and panel (b) is the strong driving case with $F=10.0 \omega_\mrb$ and $g = \omega_\mrb$. In both (a) and (b), $\omega_\mrc = \omega_\mr{d} = \omega_\mrb$.}
\label{fig:figTLSsupp1}
\end{figure}
Here, as the notation suggests, the detunings $\Delta_{\mr{Cd}}=\omega_\mrc-\omega_\mr{d}$ and $\Delta_{\mr{Bd}}=\omega_\mrb-\omega_\mr{d}$ are with respect to the driving frequency. The above equations essentially result from a transformation of the Schr\"{o}dinger picture master equation [Eq.~(1) in the main paper] with the unitary 
\begin{align}
\hat{U}_{\mr{d}}(t)=\exp(-i\omega_\mr{d} [\hsp_{\mr{C}}\hsm_{\mr{C}} + \hsp_{\mr{B}}\hsm_{\mr{B}}]t), \label{eq:UdTLS}
\end{align}
instead of the transformation to the interaction picture employed in the main paper. In this frame rotating with the drive frequency, the Hamiltonian reads:
\begin{align}
\olsi{\Hop} =&\ \Delta_\mr{Cd} \hs_\mrc^+ \hs_\mrc^- + \Delta_\mr{Bd} \hs_\mrb^+ \hs_\mrb^- + g\left(\hs_\mrc^+ \hs_\mrb^- + \hs_\mrc^- \hs_\mrb^+\right) \nonumber\\ 
    &+ F(\hs_\mrc^+ + \hs_\mrc^- ).
\label{eq:TwoTLSHamiltonianEffective}
\end{align}
After breaking up the complex expectation values like $\avg{\hs^-_{\mr{C}}}$ into their real and imaginary parts,  Eqs.~\eqref{eq:TwoTLSEOMv2set1} and \eqref{eq:TwoTLSEOMv2set2} reduce to simple homogeneous systems of first order ODEs with time-independent coefficients of the succinct form:
\begin{align}
    \frac{d \Vec{V}_1(t)}{dt} &= \mathbf{M}_1 \Vec{V}_1(t) 
    \label{eq:TwoTLSvecEOM1},\\
    \frac{d \Vec{V}_2(t)}{dt} &= \mathbf{M}_2 \Vec{V}_2(t) 
    \label{eq:TwoTLSvecEOM2}.
\end{align} 
Here, the vectors $\Vec{V}_1$ and $\Vec{V}_2$ are $\Vec{V}_1 = (\splitatcommas{\avg{\hs^z_{\mr{B}}},\avg{\hs^z_{\mr{C}}},\mathfrak{Re}[\avg{\hs^+_\mrc \hs^-_\mrb}],\mathfrak{Im}[\avg{\hs^+_\mrc \hs^-_\mrb}], \mathfrak{Re}[\avg{\hs^z_\mrc \hs^-_\mrb}] , \mathfrak{Im}[\avg{\hs^z_\mrc \hs^-_\mrb}],\mathfrak{Re}[\avg{\hs^-_\mrc \hs^-_\mrb}], \mathfrak{Im}[\avg{\hs^-_\mrc \hs^-_\mrb}],\mathfrak{Re}[\avg{\hs^-_\mrc}],\mathfrak{Im}[\avg{\hs^-_\mrc}]})^T$ and $\Vec{V}_2 = (\splitatcommas{\mathfrak{Re}[\avg{\hs^-_\mrb}],\mathfrak{Im}[\avg{\hs^-_\mrb}],\mathfrak{Re}[\avg{\hs^-_\mrc \hs^z_\mrb}],\mathfrak{Im}[\avg{\hs^-_\mrc \hs^z_\mrb}],\avg{\hs^z_\mrc \hs^z_\mrb}})^T$, and $\mathbf{M}_1,\ \mathbf{M}_2$ are $10 \times 10,\ 5\times 5$ (respectively) matrices that can be read off from Eqs.~\eqref{eq:TwoTLSEOMv2set1} and \eqref{eq:TwoTLSEOMv2set2}. 
\begin{figure}
\subfloat{\begin{overpic}[width=0.85 \linewidth]{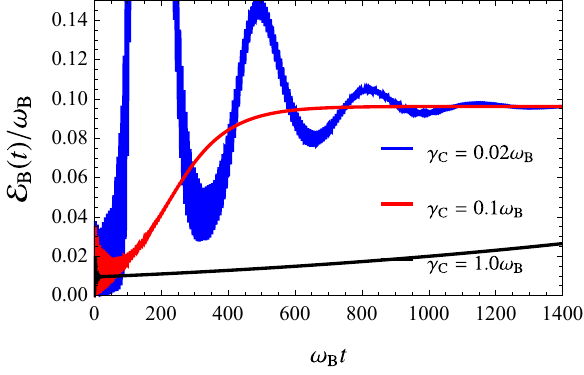}
	\put(80,58){\textbf{(a)}}
	\end{overpic}
	}\\
\subfloat{\begin{overpic}[width=0.85 \linewidth]{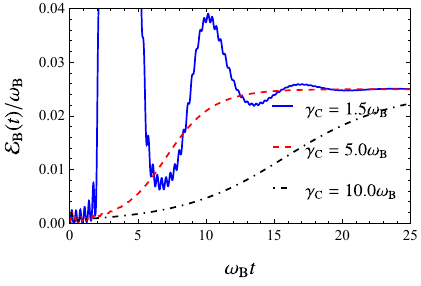}
	\put(80,58){\textbf{(b)}}
	\end{overpic}
	}
 \caption{(Color Online) Time evolution of the ergotropy $\mathcal{E}_{\mr{B}}(t)$ of the battery for two-TLS model for different values of the dephasing rate $\gamma_{\mr{C}}$. Panel (a) is for the weak driving case with parameters $F=0.1 \omega_\mrb$ and $g = \omega_\mrb$, and panel (b) is for the strong driving case with $F=10.0 \omega_\mrb$ and $g = \omega_\mrb$. In both (a) and (b), $\omega_\mrc = \omega_\mr{d} = \omega_\mrb$.
}
\label{fig:figTLSsupp2}
\end{figure}

\subsection{Resonant Case --- Analytical Solution}
\label{app:A1}
\begin{figure}
	\centering
	\subfloat{\begin{overpic}[width=0.9\linewidth]{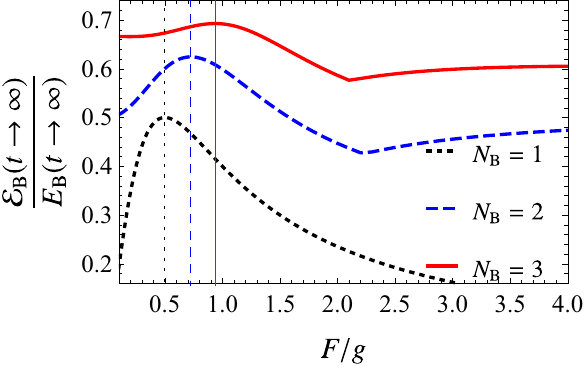}
	\end{overpic}
	}\\ 
 \caption{(Color Online) Enhancement of the steady-state ergotropy relative to the total energy by collective coupling. Asymptotic steady-state values of ergotropy of $N$ TLS batteries in a star configuration coupled to a single driven TLS charger system as a function of driving strength $F/g$ for the resonant case, i.e. $\omega_\mrc = \omega_\mr{d} = \omega_\mrb$.}
    \label{fig:StartConfigSS}
\end{figure}
Considering the resonant case with $\Delta_{\mr{CB}} = \Delta_{\mr{Cd}} = 0$, Eqs.~\eqref{eq:TwoTLSvecEOM1} and \eqref{eq:TwoTLSvecEOM2}, with the initial condition $\Vec{V}_1(0) = (\splitatcommas{-1, -1, 0, 0, 0, 0, 0, 0, 0, 0})^T$ and $\Vec{V}_2(0) = (\splitatcommas{0, 0, 0, 0, 1})^T$,
can be solved by taking the Laplace transform $\mathcal{L}[\Vec{V}_i(t)] = \int_0^{\infty}dt\, e^{-st}\Vec{V}_i(t)$ on both sides to obtain,
\begin{align}
   \mathcal{L}[\Vec{V}_i(t)] &= \frac{1}{\left(s\mathcal{I} - \mathbf{M}_i\right)^{-1}}\Vec{V}_i(0),
    \label{app:TLSeq5}
\end{align}
with $i=1,2$. From the inverse Laplace transform of the above relation, we immediately obtain the following expressions for $\avg{\hsz_{\mrb}}$ and $\avg{\hsp_{\mrb}}$, which are the variables of interest to compute the energy and ergotropy:
\begin{align}
      \avg{\hs^z_{\mrb}}(t) =& -\frac{e^{-\frac{\gamma_\mrc}{4}t}}{2\left(1+4\frac{F^2}{g^2}\right)} \Bigg\{ \frac{8F^2}{g^2} \chi_t(\gamma_\mrc,g,f_0) \nonumber \\
      & \left . +  \left(1+\sqrt{1+\frac{4F^2}{g^2}}\right) \chi_t(\gamma_\mrc,g,f_1) \right . \nonumber \\
      & + \left(1-\sqrt{1+\frac{4F^2}{g^2}}\right) \chi_t(\gamma_\mrc,g,f_2)
    \Bigg\}
    \label{app:TLSeq6}
\end{align}
and
\begin{align}
      &\avg{\hsm_{\mrb}}(t) = \frac{-\frac{F}{g}}{1+\frac{4F^2}{g^2}} + e^{-\frac{\gamma_\mrc t}{4}} \Bigg\{ \cosh\left[\frac{t}{4}\sqrt{\gamma_\mrc^2-16\left(g^2+4F^2\right)} \right] \nonumber\\
      &\quad + \frac{\gamma_\mrc}{\sqrt{\gamma_\mrc^2-16\left(g^2+4F^2\right)}}\sinh\left[\frac{t}{4}\sqrt{\gamma_\mrc^2-16\left(g^2+4F^2\right)}\right] \Bigg\}
    \label{app:TLSeq7},
\end{align}
with the functions $f_i$ and $\chi_t(\gamma_\mrc,g,f_i)$ defined in Appendix A of the main paper. From Eqs.~\eqref{app:TLSeq6} and \eqref{app:TLSeq7}, we can write down the exact expressions for the average energy $E_{\mrb}(t)$ (given in Appendix A of the main paper) and ergotropy $\mathcal{E}_{\mrb}(t)$ of the battery using Eqs.~(4) and (5) of the main paper. Note that we have plotted the battery energy for small value (weak driving) of $F/g = 0.1$ and large value (strong driving) of $F/g = 10$ in Figs.~\ref{fig:figTLSsupp1}(a) and \ref{fig:figTLSsupp1}(b), respectively. In both cases, we find that a moderate amount of dephasing leads to fast charging underscoring the generality of the central result presented in the main paper. Coming to the ergotropy, as mentioned in the main paper, the exact analytical expression is cumbersome and we do not present it here for brevity. Nonetheless, we have evaluated this expression and plotted the results for moderate driving in Fig.~2(b) of the main paper and in Figs.~\ref{fig:figTLSsupp2}(a) and \ref{fig:figTLSsupp2}(b) for weak and strong driving, respectively. The ergotropy has the same qualitative behaviour as the energy. 

Let us now consider the steady-state solution of Eqs.~\eqref{eq:TwoTLSvecEOM1} and \eqref{eq:TwoTLSvecEOM2}. For the resonant case, we have
\begin{align*}
    \mr{det}(\mathbf{M}_1) &= 4 F^4 g^2 \gamma_\mrc^2 (2 F^2 + g^2),\\
    \mr{det}(\mathbf{M}_2) &= 0.
\end{align*}
Thus the steady-state solution for Eq.~(\ref{eq:TwoTLSvecEOM1}) is $\Vec{V}_1(t \rightarrow \infty) = \Vec{V}_1^{\mr{ss}} = 0$. This implies $\avg{\hs^z_\mrb}(t \rightarrow \infty) = 0$, and hence we have 
\begin{align}
    E_\mrb(t \rightarrow \infty) &= \frac{\omega_\mrb}{2} \label{eq:TLSresEss}.
\end{align}
Since the determinant of $\mathbf{M}_2$ is zero, the variables in $\Vec{V}_2$ do not have a unique steady state. Nonetheless, we can take the long-time limit of the solutions with the initial conditions in question. To this end, since the ergotropy (Eq.~(5) in the main paper) depends on $\avg{\hs^-_\mrb}$ in addition to $\avg{\hs^z_\mrb}$, from Eq.\eqref{app:TLSeq7} we have that $\avg{\hs_\mrb^-}(t\rightarrow \infty) = \frac{-F/g}{1+4F^2/g^2}$ leading to:
\begin{align}
    \mathcal{E}_{\mrb}(t\rightarrow\infty) &= \frac{F/g}{1+4F^2/g^2} \omega_\mrb.
    \label{eq:TLSresErgss}
\end{align}

\modred{As mentioned in the main paper, while we have restricted our attention to conventional figures of merit like energy and ergotropy of the battery in our study, due to the bipartite and open nature of the charger-battery system, the battery reaches a state with non-zero von Neumann entropy. In general, we find that the von Neumann entropy also behaves in an oscillatory manner and increases with time and settles down in the long-time limit to some value below its maximal value (that occurs for a completely mixed state). A general feature that we find is also that the scenario with larger ergotropy corresponds to smaller entropy. In fact, for the TLS battery case we can exactly relate the entropy and ergotropy of the battery as follows. Writing the reduced density matrix of the battery in the general form,
\begin{align*}
    \hrho_\mrb = \frac{1}{2}\left(\hat{I} + \mathbf{n}\cdot \Vec{\sigma}\right)
\end{align*}
with $\mathbf{n} = \left(\avg{\hat{\sigma}^x_\mrb},\avg{\hat{\sigma}^y_\mrb},\avg{\hat{\sigma}^z_\mrb}\right)$, gives the von Neumann entropy as
\begin{align*}
    S_\mrb = -\frac{\left(1-n\right)}{2}\log\frac{\left(1-n\right)}{2} - \frac{\left(1+n\right)}{2}\log\frac{\left(1+n\right)}{2}
\end{align*}
with $n = |\mathbf{n}|$. Since we can easily relate the energy and ergotropy to the vector $\mathbf{n}$
via $E_\mrb/\omega_\mrb = (1 + n_z)/2$ and $\mathcal{E}_\mrb/\omega_\mrb = (n+n_z)/2$~\cite{Farina2019}, we can rewrite the ergotropy in the form
\begin{align}
    S_\mrb &= -\frac{\left(E_\mrb-\mathcal{E}_\mrb\right)}{\omega_\mrb}\log \frac{\left(E_\mrb-\mathcal{E}_\mrb\right)}{\omega_\mrb} \nonumber \\ 
    &-\left[1-\frac{\left(E_\mrb-\mathcal{E}_\mrb\right)}{\omega_\mrb}\right]\log\left[1-\frac{\left(E_\mrb-\mathcal{E}_\mrb\right)}{\omega_\mrb}\right] \label{eq:EntropyfnErgoErg}.
\end{align}
Thus clearly for a pure state with the ergotropy equal to the energy (maximal value), the entropy vanishes. On the other hand the state with ergotropy zero satisfies $n_z = -\vert \mathbf{n} \vert$ and it has the maximum entropy for a state with a given average energy $E_\mrb$.}

Finally, as we remarked in the conclusions of the main text, instead of a single battery we can also consider the scenario of multiple identical batteries (say $N$) coupled in a star configuration (collective coupling) to a single charger. In this case, the only change is in the Hamiltonian given by Eq.~(3) of the main paper which becomes
\begin{align}
    \Hop =&\ \omega_\mrc \hsp_{\mr{C}}\hsm_{\mr{C}}+\omega_\mrb \sum_{j=1}^N \hsp_{\mr{B},j}\hsm_{\mr{B},j} + g\left(\hsp_{\mr{C}} \sum_{j=1}^N\hsm_{\mrb,j} + \mathrm{h.c.} \right)\nonumber \\
    & + F(\hsm_\mrc e^{i\omega_{\mr{d}} t} + \hsp_\mrc e^{-i\omega_{\mr{d}} t}) 
\end{align}
with $\mathrm{h.c.}$ denoting the hermitian conjugate. With this change, as displayed in Fig.~\ref{fig:StartConfigSS}, we see that the ratio of the steady-state ergotropy to the energy can be made larger by increasing $N$. More precisely, we find this ratio to be $0.5,0.62,0.69$ for $N=1,2,3$, respectively. Corresponding values of the optimal driving strength to coupling ratios $F/g$ are $0.5, 0.73, 0.94$. In a forthcoming work \cite{ShastriFuture}, we will explore in detail the charging dynamics and other features of the set-up with multiple batteries.

\begin{figure}
\subfloat{\begin{overpic}[width=0.9 \linewidth]{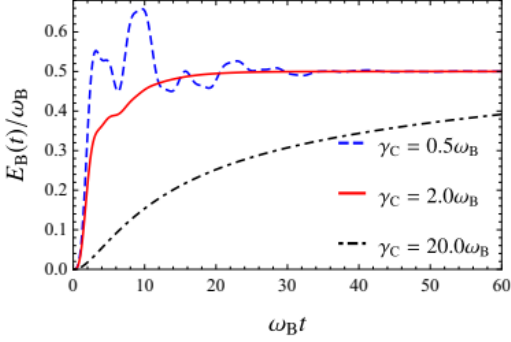}
	\put(80,58){\textbf{(a)}}
	\end{overpic}
	}\\
\subfloat{\begin{overpic}[width=0.9 \linewidth]{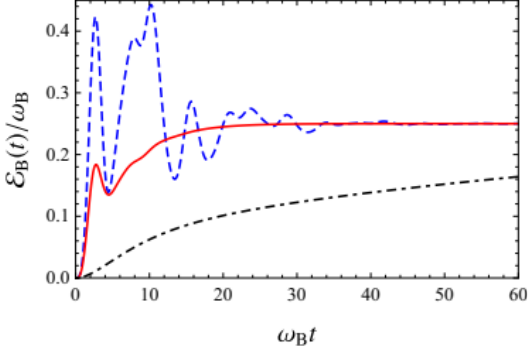}
	\put(80,58){\textbf{(b)}}
	\end{overpic}
	}
 \caption{(Color Online) Time evolution of the energy $E_{\mr{B}}(t)$ [(a)] and ergotropy $\mathcal{E}_{\mrb}(t)$ [(b)] of the battery for the two-TLS model with charger-battery detuning for different values of the dephasing rate $\gamma_{\mr{C}}$. The charger-battery detuning is set to $\Delta_{\mrc \mrb} = \Delta_{\mrc \mr{d}} = 0.02 \omega_\mrb$, and $\Delta_{\mrb \mr{d}}=0$. The driving is in the moderate regime with $F=0.5 \omega_\mrb$ and $g=\omega_\mrb$.}
\label{fig:figTLSdetEnrgErgodynCB}
\end{figure}
\subsection{Resonant Case --- Charging Time Analysis in the \texorpdfstring{$\gamma_\mrc \gg \{ F, g\}$}{gc<<{F,g}} Limit}\label{app:A2}
We now provide some additional details justifying the charging time analysis in the large dephasing limit of $\gamma_\mrc \gg \{ F,g \}$ that led to the final result presented in Eqs.~(7) and (8) in the main paper. As we discussed in Appendix B of the main paper, we begin by considering the exact expression for the energy in Eq.~(14) of the main paper, and considering the limit of $\gamma_\mrc \gg \{F,g\}$. Making the relevant approximation for the function $\chi_t(\gamma_\mrc,g,f_j)$ in this limit given in Eq.~(18) of the main paper, the condition determining the charging time [Eq.~(2) of the main paper] reduces to
\begin{align}
     \frac{1}{2\left(1+\frac{4F^2}{g^2}\right)}&\left \vert \frac{8F^2}{g^2} e^{-\frac{2g^2}{\gamma_{\mrc}}\tau} + \left(1+\sqrt{1+\frac{4F^2}{g^2}}\right)e^{-\frac{4f_1 g^2}{\gamma_{\mrc}}\tau} \right . \nonumber\\
     &\left .+\left(1-\sqrt{1+\frac{4F^2}{g^2}}\right)e^{-\frac{4 f_2\, g^2}{\gamma_{\mrc}}\tau} \right \vert = e^{-n}
  \label{eq:taugammaLarge1app}.
\end{align}
Note that the above equation was presented as Eq.~(19) of the main paper. Let us now briefly analyze the large and small driving limit of Eq.~\eqref{eq:taugammaLarge1app}. In the large driving limit with $F/g \gg 1$, we have 
$f_1 = 1-\frac{2F}{g}+2\frac{F^2}{g^2} + O(g/F)$ and $f_2 = 1+\frac{2F}{g}+2\frac{F^2}{g^2} + O(g/F)$. Thus, we have the inequality
\begin{align*}
    \frac{f_2 g^2}{\gamma_\mrc}>\frac{f_1 g^2}{\gamma_\mrc}>\frac{2 g^2}{\gamma_\mrc},
\end{align*}
which leads to the conclusion that the weakest damping scale that sets the charging time is precisely given by $\frac{2 g^2}{\gamma_\mrc}$. In the weak driving limit of $F/g \ll 1$, we have $f_1 = 2\frac{F^4}{g^4} + O(F^6/g^6)$ and $f_2 = 2 + 4 \frac{F^2}{g^2} + O(F^4/g^4)$, and the inequality
\begin{align*}
    \frac{f_2 g^2}{\gamma_\mrc} > \frac{2 g^2}{\gamma_\mrc} >\frac{f_1 g^2}{\gamma_\mrc}
\end{align*}
which leads to the conclusion that the weakest damping time scale that sets the charging time is precisely given by $\frac{f_1 g^2}{\gamma_\mrc}$. This simplifies Eq.~\eqref{eq:taugammaLarge1app} which determines the charging time to
\begin{align*}
    \frac{1+\sqrt{1+4\frac{F^2}{g^2}}}{2\left(1+\frac{4F^2}{g^2}\right)}e^{-\frac{4f_1g^2}{\gamma_\mrc}\tau} = e^{-n}.
\end{align*}
\begin{figure}
\subfloat{\begin{overpic}[width=0.85 \linewidth]{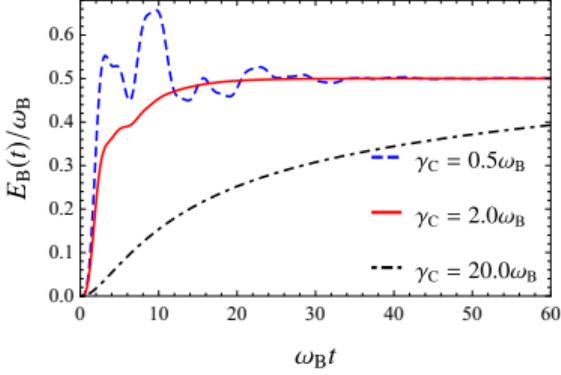}
	\put(80,58){\textbf{(a)}}
	\end{overpic}
	}\\
\subfloat{\begin{overpic}[width=0.85 \linewidth]{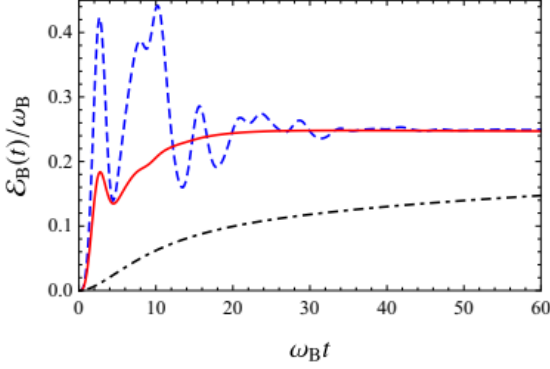}
	\put(80,58){\textbf{(b)}}
	\end{overpic}
	}
 \caption{(Color Online) Time evolution of the energy $E_{\mr{B}}(t)$ [(a)] and ergotropy $\mathcal{E}_{\mrb}(t)$ [(b)] of the battery for the two-TLS model with detuned driving for different values of the dephasing rate $\gamma_{\mr{C}}$. The driving-battery/charger detuning is set to $\Delta_{\mrb \mr{d}} = \Delta_{\mrc \mr{d}} = 0.02 \omega_\mrb$, and $\Delta_{\mrc \mrb}$ = 0. The driving is in the moderate regime with $F=0.5 \omega_\mrb$ and $g=\omega_\mrb$.}
\label{fig:figTLSdetEnrgErgodynBd}
\end{figure}
Taking the logarithm and expanding the left-hand side to lowest order in $F/g$, which turns out to be $g^4/F^4$, we find the result $\tau = \frac{ng^2}{F^4} \gamma_\mrc$ which is given as Eq.~(7) in the main paper.

We finish our consideration of the resonant case by estimating the optimal dephasing $\gamma_\mrc^\star$ by identifying the underdamped to the overdamped transition of the exact expression of the average energy $E_\mrb(t)$ given in Eq.~(14) of the main text. Let us first consider the limit of small driving $F/g \ll 1$. In the limit of $F/g \ll 1$, comparing the relative amplitude of the three contributing terms, it becomes immediately clear that the largest amplitude term is the one with $1+\sqrt{1+4 \frac{F^2}{g^2}} \approx 2$. Looking at the dynamical part of this term, it has an underdamped to overdamped transition when $\gamma_\mrc^2 = 32 f_1$. Also in the limit of small $F/g \ll1$, $f_1 \approx 2 \frac{F^4}{g^4}$, thus we have the estimate of the first transition as,
\begin{align}
\gamma_\mrc^\star \stackrel{F\ll g}{\sim} \frac{8 F^2}{g}. 
\label{gcsmallF1}    
\end{align}
For large $F/g\gg 1$ case, since $8F^2/g^2/(1+4 F^2/g^2) \approx 2$ is much larger than $(1 \pm \sqrt{1+4F^2/g^2}))/(1+4 F^2/g^2) \approx \pm 1/\sqrt{1+4F^2/g^2}$, the first term is the one with the largest magnitude and it has an underdamped to overdamped transition at 
\begin{align}
\gamma_\mrc^{\star} \sim 4  g
\label{gclargeF1}.
\end{align}
We could have also arrived approximately at the results in Eqs.~\eqref{gcsmallF1} and \eqref{gclargeF1} by equating the charging time in the small $\gamma_\mrc$ limit
[Eq.~(6) of the main paper] to the ones in the large $\gamma_\mrc$ limit [Eqs.~(7) and (8) in the main paper]. This leads to $\gamma_\mrc^\star \stackrel{F\ll g}{\approx} \frac{8 F^2}{\sqrt{2}g}$ and $\gamma_\mrc^\star \stackrel{F\gg g}{\approx} 2 \sqrt{2} g$ which agrees qualitatively with Eqs.~\eqref{gcsmallF1} and \eqref{gclargeF1}.
\subsection{Detuned Case --- Steady State and Dynamics}
\label{app:A3}
\begin{figure}
	\includegraphics[width=0.85 \linewidth]{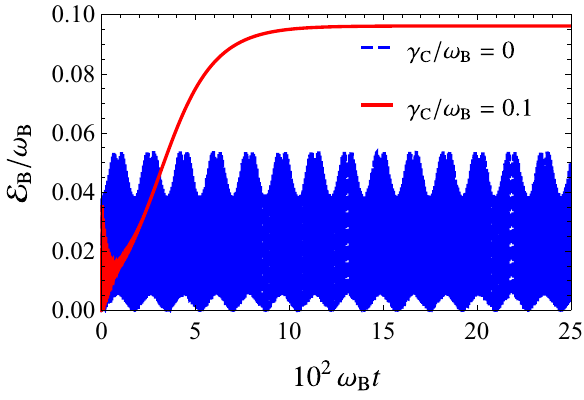}
    \caption{Transient dynamics of the ergotropy $\mathcal{E}_{\mrb}(t)$ for the two-TLS model with charger-battery detuning. Here we take weak driving regime at $F/g=0.1$ and set the detunings $\Delta_{\mr{CB}}=\Delta_{\mr{Cd}}=0.03 \omega_{\mrb}$, and $\Delta_{\mr{Bd}} = 0$. The blue dashed line is the case without charger dephasing $\gamma_\mrc = 0$ and the red solid line is with $\gamma_\mrc = 0.1 \omega_\mrb$.}
    \label{fig:detTLSergodynCB}
\end{figure}
Let us now consider the case with the detunings $\Delta_{\mr{Bd}},\Delta_{\mr{Cd}}$ non-zero. While in principle we can solve the equations of motion ~\eqref{eq:TwoTLSvecEOM1} and ~\eqref{eq:TwoTLSvecEOM2} analytically in this case, we find that in general the expressions for these solutions, both with and without dephasing $\gamma_\mrc$, are unwieldy and do not add insight. Nevertheless, as we show below, we can make some strong statements regarding the long-time or steady-state behavior of the figures of merit in this case (with $\gamma_\mrc \neq 0$). 

We will consider two kinds of detuning --- (i)~the battery-charger detuned from each other and the charger driving resonant with the battery, i.e., $\Delta_{\mr{CB}} = \Delta_{\mr{Cd}} \neq 0$ (with $\Delta_{\mrc \mrb} = \omega_\mrc - \omega_\mrb$) and $\Delta_{\mr{Bd}} = 0$ (which was discussed in the main paper); (ii) the battery and charger resonant with each other but detuned from the drive, i.e., $\Delta_{\mr{Bd}} = \Delta_{\mr{Cd}} \neq 0$, and $\Delta_{\mr{CB}} = 0$. We begin by looking at the charging dynamics in the detuned scenario. Figures~\ref{fig:figTLSdetEnrgErgodynCB} and \ref{fig:figTLSdetEnrgErgodynBd} show the dynamics of energy 
and ergotropy for the battery-charger detuning and detuned driving scenarios, respectively. As we remarked in the main paper, our central result that moderate dephasing leads to fast charging holds with detuning as well. We have chosen a moderate driving strength scenario of $F/g = 0.5$ in Figs.~\ref{fig:figTLSdetEnrgErgodynCB} and \ref{fig:figTLSdetEnrgErgodynBd} as an example, but our results apply to any value of $F$ and $g$.

\begin{figure}
\includegraphics[width=0.85 \linewidth]{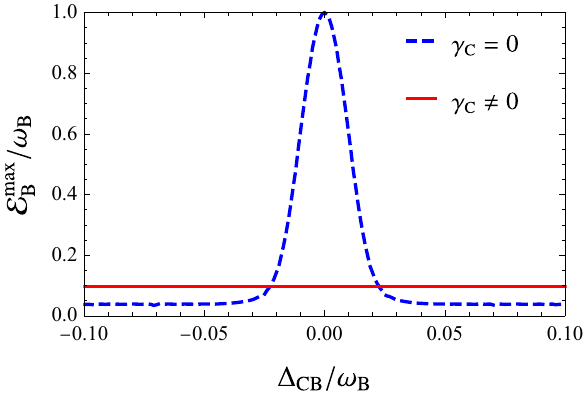}
    \caption{Maximum value of ergotropy as a function of charger-battery detuning for the two-TLS model. The solid red line represents the steady state value of the ergotropy with non-zero dephasing $\gamma_\mrc$ and the dashed blue line represents the case with no dephasing. Other parameters are $F=0.1\omega_\mrb$ and $g = 1.0 \omega_\mrb$.}
\label{fig:detTLSergoCB}
\end{figure}

Next, we would like to compare the dynamics with detuning in the presence and absence of dephasing. Considering the battery-charger detuning case first, first note that as illustrated in Fig. 4(a) of the main paper and in Fig.~\ref{fig:detTLSergodynCB}, the energy and ergotropy are oscillatory in the case with $\gamma_\mrc = 0$. In contrast, for the case with dephasing the determinants of $\mathbf{M}_i$ in Eqs.~\eqref{eq:TwoTLSvecEOM1} and \eqref{eq:TwoTLSvecEOM2} reduce to:
\begin{align*}
    \mr{det}(\mathbf{M}_1)&=4F^4g^2\gamma_\mrc\left(g^2\gamma_\mrc + 2 F^2(\gamma_\mrc-2\Delta_{\mr{CB}}) \right), \\
    \mr{det}(\mathbf{M}_2)&=0. 
\end{align*}
As a result, the steady solution of Eq.~\eqref{eq:TwoTLSvecEOM1} becomes $\Vec{V}_1 = 0$ and steady-state energy of $E_\mrb(t\rightarrow \infty) = \omega_\mrb/2$. In contrast, as before, the steady-state value of ergotropy has to be determined by calculating $\avg{\hs_\mrb^-}(t)$ and taking the long-time limit. When we do this, we find that the ergotropy is given by 
\begin{align}
    \mathcal{E}_{\mrb}(t\rightarrow\infty) &= \frac{F/g}{1+4F^2/g^2} \omega_\mrb,\label{eq:ergowithdetTLSlongtime}
\end{align} %
which is the same result that we found in the resonant case. Comparing the behaviour in the cases with and without dephasing, we showed in the main paper that the steady state energy in the case with dephasing can be larger than the maximum energy that can be attained in the closed case. In Fig.~\ref{fig:detTLSergodynCB} we show that for the same parameters as in Fig.~4(a) of the main paper the ergotropy also reaches steady-state values larger than the maximum in the closed case. This robustness provided by dephasing is summarized in Fig.~\ref{fig:detTLSergoCB} where we see that the ergotropy in the case with dephasing can be larger than the maximum in the closed case except for very small $\vert \Delta_{\mrc \mrb} \vert$ near resonance.

\begin{figure}
\includegraphics[width=0.9 \linewidth]{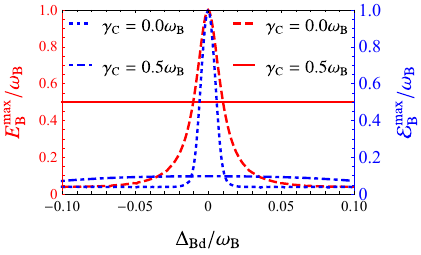}
    \caption{Maximum value of ergotropy and
average energy as a function of drive-battery detuning $\Delta_{\mrb \mr{d}}$. The long-dashed and solid red lines (left vertical axis) and the small-dashed and dash-dotted blue lines (right vertical axis) represent the maximum energy and
ergotropy, respectively. Other parameters are $F=0.1\omega_\mrb$ and $g = 1.0 \omega_\mrb$.}
\label{fig:detTLSergoenrgBd}
\end{figure}

Considering the case of the charger driving detuned from the battery and charger frequencies, we have that the determinants of $\mathbf{M}_i$,
\begin{align*}
    &\mr{det}(\mathbf{M}_1) = \frac{F^2 g^2 \gamma_\mrc}{2} \bigg\{ 16 F^4(\gamma_\mrc - 4\Delta_{\mr{Cd}}) + 4 F^2 \Big[2g^2\gamma_\mrc \\
    &\qquad  + (\gamma_\mrc-4\Delta_{\mr{Cd}})^2\Delta_{\mr{Cd}} \Big]+\gamma_\mrc\Delta_{\mr{Cd}}^2(\gamma_\mrc^2+16\Delta_{\mr{Cd}}^2) \bigg\} ,\\
    &\mr{det}(\mathbf{M}_2) = -2F^2\gamma_{\mrc}\Delta_{\mr{Cd}}^2,
\end{align*}
are non-zero. As a result, the steady solutions of Eqs.~\eqref{eq:TwoTLSvecEOM1} and \eqref{eq:TwoTLSvecEOM2} are given by $\Vec{V}_1 = \Vec{V}_2 = 0$. This immediately means that for this case, we have the steady-state energy and ergotropy take the values
\begin{align*}
    E_\mrb(t\rightarrow \infty) &= \frac{\omega_\mrb}{2},\\
    \mathcal{E}_\mrb(t\rightarrow \infty) &= 0.
\end{align*}
Interestingly, as evident from Fig.~\ref{fig:figTLSdetEnrgErgodynBd}(b), in this case the ergotropy attains a very long-lived quasi-steady state value before ultimately decaying to zero. Moreover, this quasi-steady state value of the ergotropy can be larger than the oscillating ergotropy for the closed case. We illustrate this result in Fig.~\ref{fig:detTLSergoenrgBd}, where we can see that both the energy and ergotropy in the case with dephasing can be larger than the case without dephasing for a wide region of $\Delta_{\mrb \mr{d}}$ except for very small $\vert \Delta_{\mrb \mr{d}} \vert$ near resonance. Finally, we note that our result that dephasing the charger provides robustness against the detuning continues to hold for the battery energy for any value of $F/g$ but does not hold for the ergotropy for stronger driving, i.e., for larger values of $F/g$. 
\section{HO Battery-Charger Model}
We now consider a set-up with both the charger and battery consisting of quantum harmonic oscillators (HOs). What follows will both reinforce the generality of the central result regarding moderate dephasing leading to fast charging presented in the main paper as well as highlight some expected but important differences between the HO and TLS systems. 
The Hamiltonian for the HO charger and battery is given by
\begin{align}
\Hop =&\ \omega_\mrc \adop_{\mr{C}}\aop_{\mr{C}}+\omega_\mrb \adop_{\mr{B}}\aop_{\mr{B}} + g(\adop_{\mr{C}}\aop_\mrb + \adop_\mrb\aop_\mrc)\nonumber \\
& + F(\aop_\mrc e^{i\omega_{\mr{d}} t} + \adop_\mrc e^{-i\omega_{\mr{d}} t}) \label{eq:bareHO},
\end{align}
with $\aop_\mrb$ and $\aop_\mrc$ denoting the annihilation operators for the battery and charger HO, respectively. The jump operator in the master equation~(1) of the main paper for the charger dephasing is given by $\Lop_\mrc = \adop_\mrc \aop_\mrc$. Note that in both the TLS and HO case, the choice of the coupling is such that $\commu{\Hop_{\mr{C}}+\Hop_{\mr{B}}}{\Hop_{\mr{CB}}}=0$ ensuring that there is no energetic cost to switch on/off the interaction in the absence of the charger driving with the battery and charger at resonance. 
Similar to the TLS case, we first transform to an interaction picture with respect to the bare Hamiltonians of the charger and battery with the unitary transformation $\hat{U}_{\mr{CB}}=\exp(-i[\omega_\mrc \Copd \Cop+\omega_\mrb \Bopd \Bop]t)$ applied on the bare Hamiltonian given by Eq.~\eqref{eq:bareHO} resulting in the following Hamiltonian
\begin{align}
    \Hop^\prime =&\ g\left( \Copd \Bop e^{i \Delta_{\mr{CB}}t}+ \Bopd \Cop e^{-i \Delta_{\mr{CB}} t} \right) \nonumber\\ 
    & + F(\Cop e^{-i\Delta_{\mr{Cd}} t} + \Copd e^{i\Delta_{\mr{Cd}} t}) \label{eq:HintHO},
\end{align}
with the detunings defined as in the TLS case. As before, since the jump operator $\Lop_\mrc = \Copd \Cop$ is invariant under the unitary, the time-evolution generated by Eq.~(1) of the main paper is modified to
\begin{align}
     &\frac{d\hrho^\prime(t)}{dt} = -i\commu{\Hop^{\prime}}{\hrho^\prime(t)} \nonumber \\ 
     &\quad + \frac{\gamma_\mrc}{2} \left(\Copd \Cop\, \hrho^\prime(t)\, \Copd \Cop - \frac{\{(\Copd \Cop)^2,\hrho^{\prime}(t)\}}{2}\right)
    \label{eq:HOmasterinteraction},
\end{align}
with $\hrho^\prime = \hat{U}_{\mr{CB}}^{\dagger}\, \hrho\, \hat{U}_{\mr{CB}}$. While the total energy stored in the battery can be written as
\begin{figure}
	\centering
	\subfloat{\begin{overpic}[width=0.8\linewidth]{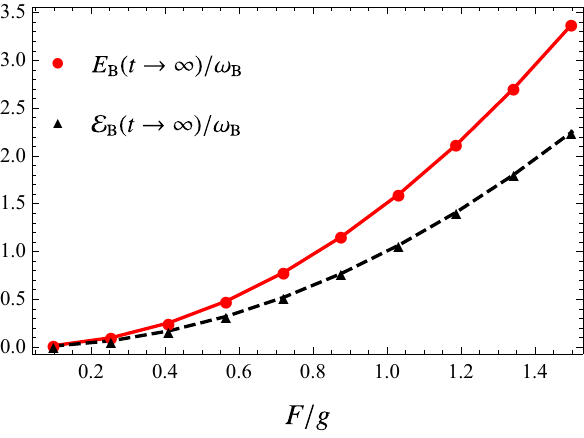}
	\end{overpic}
	}\\ 
  \caption{(Color Online) 
  Asymptotic steady-state values of average energy $E_{\mr{B}}(\infty)$ and ergotropy $\mathcal{E}_{\mr{B}}(\infty)$ of the HO battery-charger set-up with dephasing. Here, solid red line (with dot markers) represents the analytical values, and the dashed black line is the fitted values $\left(F/g\right)^2$ to $\mathcal{E}_{\mr{B}}(\infty)$. Steady state values of energy and ergotropy are dependent only on $F/g$ and are independent of $\gamma_\mrc$.}
    \label{fig:HOfig3}
\end{figure}
\begin{align}
    E_{\mr{B}}(t) = \omega_\mrb \avg{\Bopd \Bop} = \omega_\mrb \Tr{\hrho^\prime(t)\, \Bopd \Bop}{\mr{BC}}, \label{eq:engHO}
\end{align}
for arbitrary (possibly non-Gaussian) quantum states of the battery, a closed-form expression for the ergotropy in terms of expectations of battery operators is not derivable \cite{Farina2019,Francica2020}. Indeed in contrast to the work of \cite{Farina2019}, where the dissipative evolution was Gaussian, the dephasing jump operator generates non-Gaussian states of the oscillators \cite{Francica2020}. Thus, we need to resort to numerical solutions of the master equation to compute the ergotropy. 
Such numerical simulations in general will involve a cut-off in the number of oscillator eigenstates and the required number of states to capture the dynamics faithfully will increase with the ratio between the driving and coupling strength $F/g$. Interestingly, as we discuss next in the oscillator case, for resonant driving there is no qualitative distinction in the dynamics for different values of the ratio between the driving and coupling strength $F/g$ in the sense of its dependence on the charger dephasing rate $\gamma_\mrc$. Moreover, we will show that there is a simple scaling relation connecting the dynamics for different $F/g$.
\begin{figure}
	\centering
	\subfloat{\begin{overpic}[width=0.8 \linewidth]{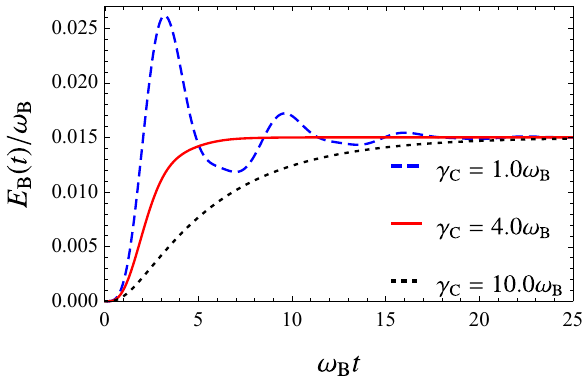}
	\put(80,55){\textbf{(a)}}
	\end{overpic}
	}\\
	\subfloat{\begin{overpic}[width=0.8 \linewidth]{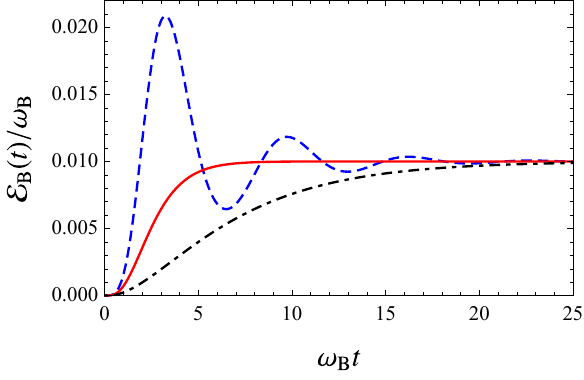}
	\put(80,55){\textbf{(b)}}
	\end{overpic}
	}\\
    \caption{(Color Online) Time Evolution of average energy $E_{\mr{B}}(t)$ [(a)] and ergotropy $\mathcal{E}_{\mr{B}}(t)$ [(b)] for the HO battery-charger set-up. Ratio of driving to coupling is $F/g=0.1$ and the dephasing $\gamma_{\mr{C}}$ is varied. Other parameter values are $g=1.0\omega_\mrb$, and $\omega_\mrc = \omega_\mr{d} = \omega_\mrb$.}
    \label{fig:HOfig1}
\end{figure}
\subsection{Resonant Case --- Analytical Solution}
Let us, as before, first focus on the resonant case with $\Delta_{\mr{CB}} = \Delta_{\mr{Cd}} = 0$. We now perform an additional unitary transformation generated by the displacement operator $\Uop = \Dop(F/g) = \exp[F/g\, (\Bopd + \Bop)]$ which reduces the master equation~\eqref{eq:HintHO} to the following form for the transformed density operator $\hat{\tilde{\rho}} = \Uop \hrho^\prime \Uopd$:
\begin{align}
   \frac{d \hat{\tilde{\rho}}}{dt} =& -i\commu{g\left(\Copd\Bop  + \Cop\Bopd\right)}{\hat{\tilde{\rho}}(t)} \nonumber\\
   &+\gamma_{\mr{C}}\left(\Copd \Cop \hat{\tilde{\rho}}(t) \Copd \Cop - \frac{1}{2}\acommu{\Copd\Cop}{\hat{\tilde{\rho}}(t)}\right).
\label{twoHO:eq3}
\end{align}
Here, we have used the transformation $\hat{\bar{H}} = \Uop \Hop^\prime \Uopd$ with the property $\Dop(\alpha)\, \Bop\, \Dopd(\alpha) = \Bop - \alpha$ and $[\Dop(F/g),\Copd] = 0$. From Eq.~\eqref{twoHO:eq3}, we can write down the following closed set of EOMs for the moments
\begin{equation}
\begin{split}
          \frac{d\avg{\Cop}}{dt} &= -ig\avg{\Bop} - \frac{\gamma_{\mr{C}}}{2}\avg{\Cop},\\
          \frac{d\avg{\Bop}}{dt} &= -ig\avg{\Cop},\\
          \frac{d\avg{\Copd\Cop}}{dt} &= 2g\mathfrak{Im}\left[\avg{\Copd\Bop}\right],\\
          \frac{d\avg{\Copd\Bop}}{dt} &= -ig\left(\avg{\Copd\Cop} - \avg{\Bopd\Bop} \right) - \frac{\gamma_{\mr{C}}}{2}\avg{\Copd \Bop},\\
          \frac{d\avg{\Bopd\Bop}}{dt} &= -2g\mathfrak{Im}\left[\avg{\Copd\Bop}\right],
\end{split}
\label{twoHO:eq4}
\end{equation}
where the expectation values should be understood as taken with respect to the transformed density operator, i.e., $\avg{\cdot} = \Tr{\hat{\tilde{\rho}}\,\, \cdot}{\mrb \mrc}$. This holds for the initial conditions as well with the initial state $\hrho^\prime(0) = \ketbra{\psi_0}{\psi_0}$ with $\ket{\psi_0} = \ket{0}_{\mr{C}}\otimes\ket{0}_\mrb$ transformed to $\hat{\tilde{\rho}}(0) = \ketbra{\tilde{\psi}_0}{\tilde{\psi}_0}$ with $\ket{\tilde{\psi}(0)}= \ket{0}_{\mr{C}}\otimes\ket{\alpha = F/g}_\mrb$. Here $\ket{\alpha = F/g}_\mrb = \Dop(F/g) \ket{0}_\mrb$ is a coherent state of the battery.  With this, the initial conditions for solving Eq.~\eqref{twoHO:eq4} becomes $\avg{\Cop}(0) = \avg{\Copd\Cop} (0) = \avg{\Copd\Bop}(0) = 0$, $\avg{\Bop} (0) = F/g$, and $\avg{\Bopd\Bop}(0) = F^2/g^2$. We have now reformulated the dynamics such that all of the dependence on the ratio between the driving and coupling strength $F/g$ comes via the initial conditions. Solving equations \eqref{twoHO:eq4} with the above initial condition, we get the average energy of the battery with respect to the density matrix $\hrho^\prime$ as $E_{\mr{B}}(t)/\omega_\mrb = \Tr{\hrho^\prime(t)\, \Bopd\Bop}{\mr{BC}} = \Tr{\hat{\tilde{\rho}}(t)\,\Dop(F/g)\,\Bopd\Bop\,\Dopd(F/g)}{\mr{BC}}$ as,
\begin{align}
          &\frac{E_{\mr{B}}(t)}{\omega_\mrb} = \avg{\Bopd \Bop} - \frac{F}{g} \avg{\Bopd+\Bop} + \frac{F^2}{g^2} \nonumber \\
          &= \frac{F^2}{g^2}\left\{ \frac{3}{2} - \frac{e^{-\frac{\gamma_{\mr{C}}}{4}t}}{2}\left[ 4\cosh{\left(\frac{\Gamma t}{4}\right)} + 4\frac{\gamma_{\mr{C}}\sinh{\left(\frac{\Gamma t}{4}\right)}}{\Gamma}\right . \right . \nonumber\\ 
          &\left . \left . - \cosh{\left(\frac{\sqrt{\Gamma^2 - 3(4g)^2}t}{4}\right)} - \frac{\gamma_{\mr{C}}\sinh{\left(\frac{\sqrt{\Gamma^2 - 3(4g)^2}t}{4}\right)}}{\sqrt{\Gamma^2-3(4g)^2}} \right]  \right\},
          \label{twoHO:eq5}
\end{align}
with $\Gamma^2 = \gamma_{\mr{C}}^2 - (4g)^2$ (and $\Gamma$ should be read as a possible complex variable). From the above equation, it is now apparent that the ratio between the driving and coupling strength $F/g$ simply appears as a quadratic scaling factor for the energy stored in the battery. We have been able to confirm the same scaling behavior (as a function of $F/g$) for the ergotropy from numerical calculations of the dynamics. This is most easily seen from the steady state values of energy ($E_\mrb(\infty)$) and ergotropy ($\mathcal{E}_\mrb(\infty)$) displayed in Fig.~\ref{fig:HOfig3}. From Fig.~\ref{fig:HOfig3}, we see that the numerical calculations of the steady-state energy agree with the analytical expectation which can be read off from Eq.~\eqref{twoHO:eq5} as
\begin{align}
    \frac{E_\mrb(\infty)}{\omega_\mrb} = \frac{3}{2}\frac{F^2}{g^2}
    \label{twoHO:eq6},
\end{align}
and the steady state ergotropy is fitted perfectly by the expression $\mathcal{E}_\mrb(\infty)/\omega_\mrb \approx F^2/g^2$. Thus, as evident from Eqs.~\eqref{twoHO:eq5} and \eqref{twoHO:eq6} that $F/g$ merely serves as a scale for the energy and ergotropy.

\begin{figure}
	\centering
	\subfloat{\begin{overpic}[width=0.8\linewidth]{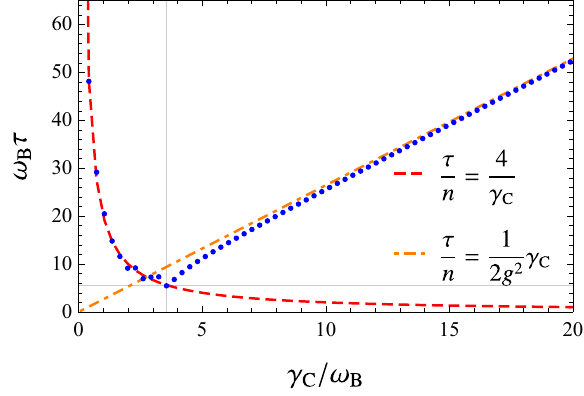}
	\end{overpic}
	}\\ 
  \caption{(Color Online) Charging time $\tau$ as a function of the dephasing rate of the charger $\gamma_{\mr{C}}$ for the two-HO model of quantum battery. Here, the red dashed line corresponds to fitting Eq.~\eqref{eq:tauweakdephasingHO}, and the orange dashed line is the fitting Eq.~\eqref{eq:taustrongdephasingHO}. The grey vertical line indicates the optimal value, $\gamma_\mrc^{\star}$, to obtain fastest charging time. Other parameter values are the same as \figref{fig:HOfig1}.}
    \label{fig:HOfig2}
\end{figure}

With this insight, let us now examine the energy and ergotropy as a function of $\gamma_\mrc$. In Fig.~\ref{fig:HOfig1}, we have plotted the dynamics of energy (a) and ergotropy (b) for three different values of $\gamma_\mrc$. In line with the consideration for the TLS, we again see that the charging time takes large values for very small or large values of $\gamma_\mrc$ with an optimum in between. Thus our central result that a moderate amount of charger dephasing leads to fast charging is valid even for the two HOs set-up. It is evident from Eq.~\eqref{twoHO:eq5} that, in the case of the HO, the charging time $\tau$ and the optimal dephasing $\gamma_\mrc^\star$ are going to be independent of the ratio between the driving and coupling strength $F/g$. As with the TLS case, let us estimate the charging time in the large and small dephasing limits using the condition given by Eq.~(2) of the main paper with $n=1$. First, note that at small dephasing $\gamma_{\mr{C}}/g \ll 1$ we get from Eq.~\eqref{twoHO:eq5} the expression,
\begin{align}
          E_\mrb(t) &\approx \frac{3}{2}\frac{F^2}{g^2} - \frac{1}{2}\frac{F^2}{g^2} e^{-\frac{\gamma_{\mr{C}}t}{4}}\left( 4\cos{gt} - \cos{2gt} \right).
        \label{twoHO:eq7}
\end{align}
From this, since there is only one damping scale, the charging time is given by 
\begin{align}
    \tau &\sim \frac{4}{\gamma_\mrc}  \label{eq:tauweakdephasingHO}. 
\end{align}
On the other hand, in the limit of large dephasing $\gamma_{\mr{C}}/g \gg 1$ we have,
\begin{align}
          E_\mrb(t) &\approx \frac{3}{2}\frac{F^2}{g^2} - 2\frac{F^2}{g^2} e^{-\frac{2g^2 t}{\gamma_{\mr{C}}}} + \frac{F^2}{2g^2} e^{-\frac{8g^2 t}{\gamma_{\mr{C}}}}.
        \label{twoHO:eq72}
\end{align}
We now have two damping time scales and we choose the weaker of the two, i.e., $\frac{2g^2 }{\gamma_{\mr{C}}}$ and thus the charging time is given by
\begin{align}
\tau &\sim \frac{1}{2g^2} \gamma_\mrc \label{eq:taustrongdephasingHO}.
\end{align}
Figure~\ref{fig:HOfig2} illustrates the charging time calculated from the analytical expression of Eq.~\eqref{twoHO:eq5} by using the condition given by Eq.~(2) of the main paper with $n=1$ and also validates the limiting behavior described by Eqs.~\eqref{eq:tauweakdephasingHO} and \eqref{eq:taustrongdephasingHO}. To obtain an accurate estimate of the optimal dephasing, let us now consider at what values of $\gamma_\mrc$ do we get underdamped to overdamped transitions in Eq.~\eqref{twoHO:eq5}. There are two such transitions possible, namely when $\Gamma^2=0$ and $\Gamma^2 - 3 (4g)^2 = 0$ which occur for $\gamma_\mrc = 4g$ and $\gamma_\mrc = 8g$, respectively. Since we expect the optimal dephasing rate to be controlled by the term with the larger amplitude of oscillation [which are the first two terms in the square bracket in Eq.~\eqref{twoHO:eq5}], we have the optimal dephasing rate as
\begin{align}
   \gamma_\mrc^\star = 4 g \label{eq:optidephtaustar_HO},
\end{align}
independent of the ratio between the driving and coupling strength $F/g$. As evident from Fig.~\ref{fig:HOfig2}, this is a very good estimate. Consequently, the fastest charging time in the HO case irrespective of driving is given by $\tau^\star \equiv 4/\gamma_\mrc^\star \approx 1/g$. As in the TLS case, we can also estimate the optimal dephasing rate by equating charging times in the small and large dephasing regimes given by Eqs.~\eqref{eq:tauweakdephasingHO} and \eqref{eq:taustrongdephasingHO}, respectively. In this manner we find the optimal dephasing rate to be $\gamma_\mrc^\star \approx 2 \sqrt{2} g$, which is very close to the result in Eq.~\eqref{eq:optidephtaustar_HO}.

In summary, we find that our central result of moderate dephasing leading to faster charging extends to the HOs case. While the optimum dephasing rate $\gamma_{\mrc}^\star$ depends on the charger driving strength and charger-battery coupling in the former, it depends solely on the charger-battery coupling in the latter. 

\subsection{\label{app:B} Detuned Case --- Dynamics and Steady State}
\begin{figure*}
	\centering
	\subfloat{\begin{overpic}[width=0.33 \linewidth]{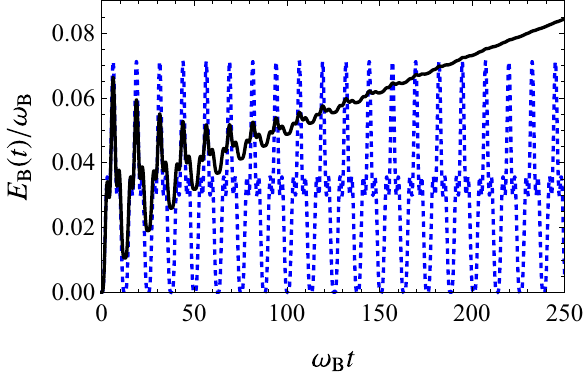}
	\put(25,55){\textbf{(a)}}
	\end{overpic}
	}
	\subfloat{\begin{overpic}[width=0.32 \linewidth]{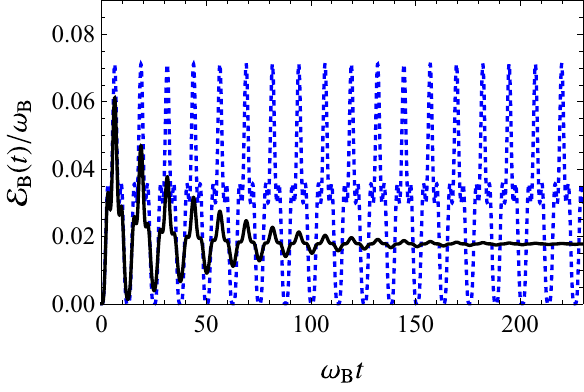}
	\put(25,55){\textbf{(b)}}
	\end{overpic}
	}
 	\subfloat{\begin{overpic}[width= 0.33\linewidth]{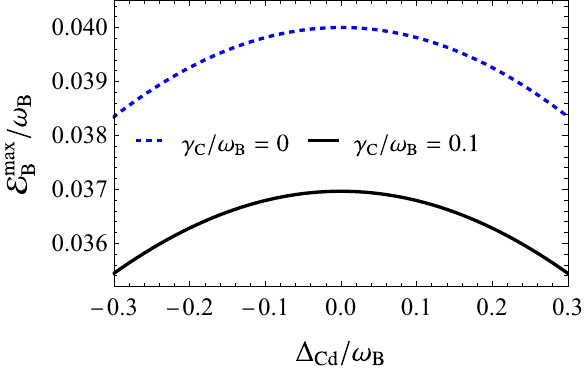}
	\put(25,55){\textbf{(c)}}
	\end{overpic}
	}
    \caption{(Color Online) Transient dynamics of the average energy $E_{\mrb}(t)$~[(a)] and ergotropy $\mathcal{E}_{\mrb}(t)$~[(b)] of the battery for two-HO model. The charger driving is detuned with $\Delta_{\mr{Cd}}=\Delta_{\mr{Bd}}=0.5 \omega_{\mr{B}}$. Maximum value of ergotropy $\mathcal{E}^{\mathrm{max}}_\mrb$~[(c)] as a function of drive detuning $\Delta_{\mr{Cd}}=\Delta_{\mr{Bd}}$. Other parameters are $F=0.1\omega_\mrb$ and $g=1.0\omega_\mrb$.}
    \label{fig:figHO4}
\end{figure*}

Let us next consider the HO battery-charger system with detuning. As with the TLS case, to discuss the analytical solutions for the two-HO model with detuning, it is convenient to work in a frame that is transformed with respect to the original master equation (1) of the main paper. We do this via the unitary transformation
\begin{align}
\Uop_{\mr{d}}(t) = e^{i\omega_\mr{d}\left(\Copd\Cop + \Bopd\Bop\right)t} \label{eq:UdHO},   
\end{align}
which is slightly different from the interaction picture transformation used in the previous sub-section while dealing with the resonant case. With this approach, the two-HO Hamiltonian in the frame rotating at the drive frequency reads
\begin{align}
\olsi{\Hop} &= \Delta_\mr{Cd} \Copd\Cop + \Delta_\mr{Bd} \Bopd\Bop + g\left(\Copd \Bop + \Bopd \Cop \right) \nonumber\\ 
    &+ F(\Cop + \Copd)
\label{eq:TwoHOHamiltonianEffective},
\end{align}
with the detunings as defined in the TLS case. The master equation in this transformed frame leads to the following equations for the operator expectation values:
\begin{align}
    \frac{d}{dt}\avg{\Bopd\Bop} =& -2g\mathfrak{Im}\left[\Copd\Bop\right], \nonumber\\
    \frac{d}{dt}\avg{\Copd\Cop} =& -2F\mathfrak{Im}\left[\avg{\Cop}\right] + 2g\mathfrak{Im}\left[\avg{\Copd\Bop}\right], \nonumber\\
    \frac{d}{dt}\avg{\Copd\Bop} =&\ iF\avg{\Bop} - ig\left(\avg{\Copd\Cop} - \avg{\Bopd\Bop}\right) \nonumber\\
    &- \left[\frac{\gamma_\mrc}{2} - i\left(\Delta_\mr{Cd}-\Delta_\mr{Bd}\right)\right]\avg{\Copd\Bop}, \nonumber\\
    \frac{d}{dt}\avg{\Bop} =& -ig\avg{\Cop} -i\Delta_\mr{Bd}\avg{\Bop},\nonumber \\
    \frac{d}{dt} \avg{\Cop} =& -iF -ig\avg{\Bop} - \left(\frac{\gamma_\mrc}{2} + i\Delta_\mr{Cd}\right)\avg{\Cop}
    \label{eq:TwoHOEOM1}.
\end{align}
\begin{figure}
	\centering
	\subfloat{\begin{overpic}[width=0.8\linewidth]{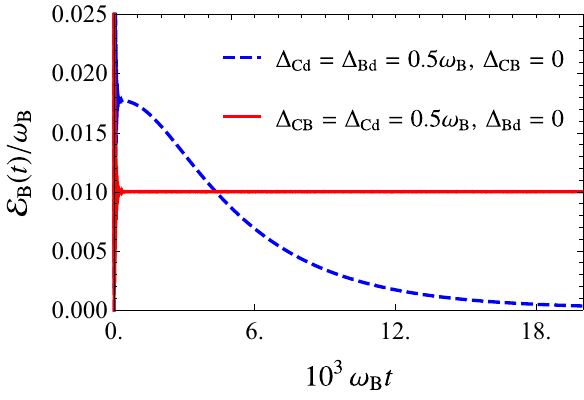}
	\end{overpic}
	}\\ 
 \caption{(Color Online) Time evolution of the ergotropy of two-HO model for the detuned driving (blue dashed line; $\Delta_\mr{CB} =0$ and $\Delta_\mr{Cd}=\Delta_\mr{Bd}=0.5\omega_\mrb$) and the detuned charger (red solid line; $\Delta_\mr{CB} = \Delta_\mr{Cd}=0.5\omega_\mrb$ and $\Delta_\mr{Bd}=0$). Other parameter values are $F=0.1\omega_\mrb$, $g=1.0\omega_\mrb$, and $\gamma_\mrc = 0.1\omega_\mrb$.}
\label{fig:HOfig1app}
\end{figure}
\begin{figure}
 \centering
	\subfloat{\begin{overpic}[width=0.8\linewidth]{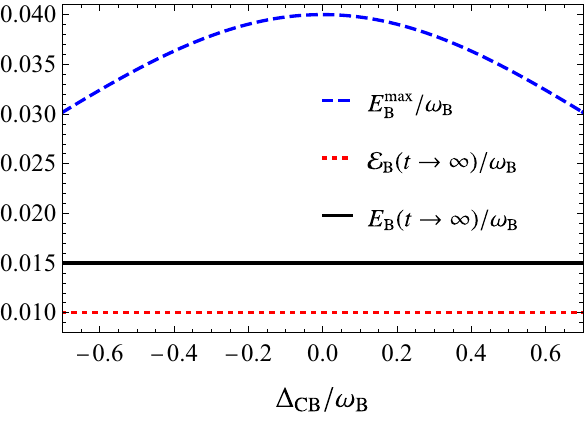}
	\end{overpic}
	}\\ 
  \caption{(Color Online) Average energy and ergotropy of two-HO model as a function of battery-charger detuning. Dashed blue line represents the maximum value of energy and ergotropy $E_\mrb^{\mr{max}}/\omega_\mrb = \mathcal{E}_\mrb^\mr{max}/\omega_\mrb$ for the zero dephasing $\gamma_\mrc=0$ case. Black solid and red short dashed lines represent the long-time steady state values of average energy and ergotropy, respectively, for non-zero dephasing $\gamma_\mrc=0.1\omega_\mrb.$ Other parameters are $F=0.1 \omega_\mrb$ and $g=1.0\omega_\mrb$.}
    \label{fig:HOfig2app}
\end{figure}
\begin{figure}
\subfloat{\begin{overpic}[width=0.8 \linewidth]{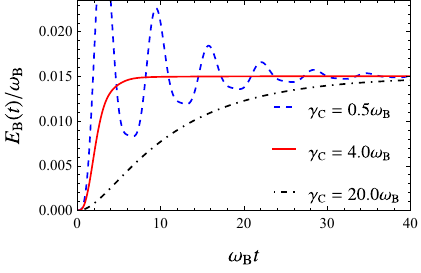}
	\put(80,58){\textbf{(a)}}
	\end{overpic}
	}\\
\subfloat{\begin{overpic}[width=0.8 \linewidth]{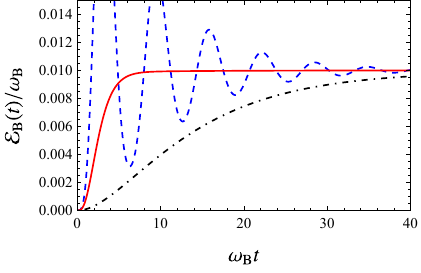}
	\put(80,58){\textbf{(b)}}
	\end{overpic}
	}
 \caption{(Color Online) Time evolution of average energy $E_{\mr{B}}(t)$~[(a)] and ergotropy $\mathcal{E}_{\mr{B}}(t)$~[(b)] for the TLS-HO battery-charger set-up. The driving has been chosen in the weak regime with $F=0.1 \omega_\mrb$ and $g = \omega_\mrb$ ($F/g=0.1$), and the dephasing $\gamma_{\mr{C}}$ is varied. Other parameter values are $g=1.0\omega_\mrb$ and $\omega_\mrc = \omega_\mr{d} = \omega_\mrb$.
}
\label{fig:figHOTLS1}
\end{figure}
\begin{figure}
\subfloat{\begin{overpic}[width=0.8 \linewidth]{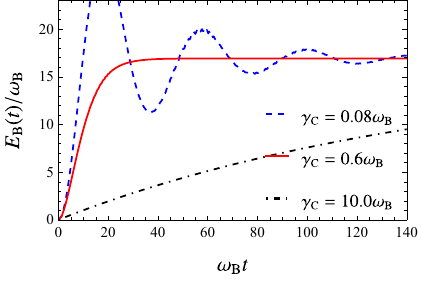}
	\put(80,58){\textbf{(a)}}
	\end{overpic}
	}\\
\subfloat{\begin{overpic}[width=0.8 \linewidth]{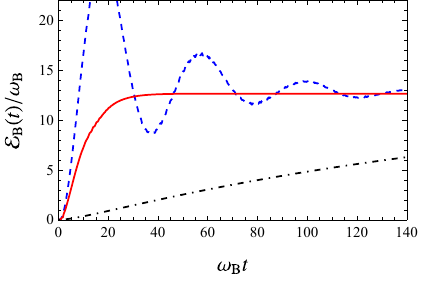}
	\put(80,58){\textbf{(b)}}
	\end{overpic}
	}
 \caption{(Color Online) Time evolution of average energy $E_{\mr{B}}(t)$~[(a)] and ergotropy $\mathcal{E}_{\mr{B}}(t)$~[(b)] for the TLS-HO battery-charger set-up. The driving has been chosen in the strong regime with $F=3.0 \omega_\mrb$ and $g = \omega_\mrb$ ($F/g=3.0$), and the dephasing $\gamma_{\mr{C}}$ is varied. Other parameter values are $g=1.0\omega_\mrb$ and $\omega_\mrc = \omega_\mr{d} = \omega_\mrb$.
}
\label{fig:figHOTLS2}
\end{figure}
As described earlier, while the solution of the above equations of motion can help us calculate the average energy of the battery, we have to solve the master equation numerically to determine the ergotropy with non-zero dephasing. Similar to the TLS case, we collect the equations for the expectation values into a matrix form as,
\begin{align}
    \frac{d \Vec{V}(t)}{dt} &= \mathbf{M}\Vec{V}(t) + \Vec{W},
    \label{eq:TwoHOEOM3}
\end{align} 
with the vectors $\Vec{V} = (\splitatcommas{\avg{\Bopd\Bop},\avg{\Copd\Cop}, \mathfrak{Im}\left[\avg{\Copd\Bop}\right], \mathfrak{Re}\left[\avg{\Copd\Bop}\right], \mathfrak{Re}\left[\Bop\right],\mathfrak{Tm}\left[\Bop\right], \mathfrak{Re}\left[\Cop\right],\mathfrak{Im}\left[\Cop\right]})^T$ and $\Vec{W} = (\splitatcommas{0,0,0,0,0,0,0,-F})^T$. Here, $\mathbf{M}$ is a $8\times 8$ matrix which can be read off from Eq.~\eqref{eq:TwoHOEOM1} after separating the real and imaginary parts. Note, in contrast to the two-TLS model, the EOMs are inhomogeneous. The density matrix at initial time $\hrho^\prime(t=0) = \ketbra{0}{0}_\mrb \otimes \ketbra{0}{0}_\mrc$ translates to the initial condition $\Vec{V}(0) = (\splitatcommas{0,0,0,0,0,0,0,0})^T$. By taking the Laplace transform $\mathcal{L}[\Vec{V}(t)] = \int_0^{\infty}dt\, e^{-st}\Vec{V}(t)$ on both sides of Eq.~\eqref{eq:TwoHOEOM3}, we obtain
\begin{align}
   \mathcal{L}[\Vec{V}(t)] &= \frac{1}{s\left(s\mathcal{I} - \mathbf{M}\right)^{-1}}\Vec{W}.
    \label{eq:TwoHOEOM4}
\end{align}
By taking the inverse Laplace transform of the above equation, we can determine $\avg{\Bopd\Bop}(t)$.

Focusing first on the zero dephasing case, we note that  Eq.~\eqref{eq:TwoHOEOM1} can be solved to obtain tractable analytical expressions. As discussed in Ref.~\cite{Farina2019}, for the closed Gaussian dynamics of the oscillators, the energy and ergotropy coincide for our choice of initial conditions. Let us first look at the case of detuned driving with $(\Delta_{\mr{Bd}}=\Delta_{\mr{Cd}})\neq 0$. By solving the equations of motion \eqref{eq:TwoHOEOM1} with non-zero detuning for the closed case, the average energy takes the form
\begin{align}
    \frac{E_\mrb(t)}{\omega_\mrb} =& \frac{F^2}{2(\Delta_{\mr{Cd}}^2-g^2)^2}\left [ 3 g^2 + g^2 \cos(2gt) \right.\nonumber\\
    & - 4 g^2 \cos(gt) \cos(\Delta_{\mr{Cd}}t) \nonumber\\ 
    & \left. + 2 \sin(gt) \Delta_{\mr{Cd}} \left(\Delta_{\mr{Cd}} \sin(gt)-2 g \sin(\Delta_{\mr{Cd}}t) \right) \right]  \label{eq:closedDetDriveHO}.
\end{align}
This purely oscillatory evolution, depicted for an exemplary parameter choice in Figs.~\ref{fig:figHO4}(a) and \ref{fig:figHO4}(b), for $\Delta_{\mr{Cd}}<g$ takes its maximum amplitude of 
\begin{align}
    \frac{E_\mrb^{\mr{Max}}}{\omega_\mrb} = \frac{4 F^2 g^2}{(\Delta_{\mr{Cd}}^2-g^2)^2} \cos^2\left ( \frac{\Delta_{\mr{Cd}\pi}}{2g}\right) \label{eq:EBmaxHOclosed},
\end{align}
at $t = (2n+1) \pi/g$. From Eqs.~\eqref{eq:closedDetDriveHO} and \eqref{eq:EBmaxHOclosed}, we can see that as $\Delta_{\mr{Cd}} \rightarrow \pm g$, there is a resonant enhancement of the battery energy with an unbounded increase of energy and ergotropy. This is expected as the normal mode frequency of the coupled battery and charger (at resonance with each other) is precisely at $\omega_\mrb \pm g$ \cite{gangwar2024coherently}. Coming to the case where the battery and charger are detuned, i.e. $\Delta_{\mr{Bd}}=0$ and $\Delta_{\mr{CB}}\neq0$, the energy is given by
\begin{align}
    &\frac{E_\mrb(t)}{\omega_\mrb} =\frac{F^2}{g^2}\left\{ 2 - \frac{2g^2}{\Delta^2_\mr{CB}+4g^2} \right. \nonumber\\
    &\qquad \left. +  \frac{2g^2\cos\left[\sqrt{\Delta^2_{\mr{CB}} + 4g^2}t\right]}{\Delta^2_\mr{CB}+4g^2 }-\left[\cos\left(\frac{gt}{\alpha}\right)+\cos\left(\alpha t\right)\right]\right. \nonumber\\
    &\qquad -\frac{\Delta_\mr{CB}}{\sqrt{\Delta^2_\mr{CB}+4g^2}}\left[\cos\left(\frac{gt}{\alpha}\right)-\cos\left(\alpha t \right)\right] 
    \Biggr\}
    \label{eq:TwoHOEOM6},
\end{align}
where $\alpha = \sqrt{g^2+\frac{1}{2}\Delta^2_\mr{CB}+\frac{1}{2}\Delta_\mr{CB}\sqrt{\Delta^2_\mr{CB}+4g^2}}$. A key point of difference with the above expression compared to Eq.~\eqref{eq:closedDetDriveHO} is that it does not diverge at $\Delta_{\mr{CB}} = \pm g$. 

As in the TLS case, for the case with dephasing and non-zero detuning, analytical solutions for the energy become involved. Nonetheless, we can extract some key features of the dynamics at long times. Considering first the case of detuned driving with $(\Delta_{\mr{Bd}}=\Delta_{\mr{Cd}})\neq 0$, we find that at long times the energy becomes a linear function of time of the form
\begin{align}
   \frac{E_\mrb(t)}{\omega_\mrb} \stackrel{\omega_{\mrb}t \gg 1}{\approx} \frac{2F^2\gamma_\mrc\Delta^2_\mr{Cd} t}{4g^4 - 8g^2\Delta^2_\mr{Cd} + \gamma_\mrc^2\Delta^2_\mr{Cd} + 4\Delta^4_\mr{Cd}}
    \label{eq:longtimeDet},
\end{align}
as shown in Fig.~\ref{fig:figHO4}(a). Thus, in contrast to the two-TLS model, looking at transient maxima of $E_\mrb(t)$ is not meaningful for the two-HO model for the detuned driving case. Coming to the ergotropy, which we calculate numerically using QuTIP \cite{Qutip}, we see in Fig.~\ref{fig:figHO4}(b) that after initial oscillations it settles to an quasi-steady value. Over much longer time scales, as shown in Fig.~\ref{fig:HOfig1app}, we find that the ergotropy damps towards zero similar to the TLS case. Let us now compare the behavior of this transient maxima of the ergotropy in the case with dephasing to the closed case. As shown in Fig.~\ref{fig:figHO4}(c), unlike the TLS case, for $\vert \Delta_{\mr{Cd}} \vert < g$ with detunings close to resonance the ergotropy with dephasing is always smaller than the closed case. While we have depicted the behavior around $\Delta_{\mr{Cd}} = 0$ to better compare with the TLS results, the ergotropy in the dephased charger case also has peaks at $\Delta_{\mr{Cd}} = \pm g$. Examining the case where the battery and charger are detuned ($\Delta_{\mrc \mrb} \neq 0, \Delta_{\mrb \mathrm{d}} = 0$), 
we find by taking the $t\rightarrow\infty$ limit from an analytical solution of Eq.~\eqref{eq:TwoHOEOM3} that the average energy becomes
\begin{align}
    \frac{E_\mrb(t\to \infty)}{\omega_\mrb} =
     \frac{3F^2}{2g^2}    
\label{eq:TwoHOEOM8},
\end{align}
which is exactly equal to the steady state in the resonant case. From a numerical solution of the master equation, as shown in Fig.~\ref{fig:HOfig1app}, we find that while the ergotropy goes to zero in the long time limit for $\Delta_\mr{Bd}=\Delta_\mr{Cd}\neq 0$ and $\Delta_{\mr{CB}}=0$, it goes to a steady non-zero value for the case of $\Delta_{\mr{CB}} = \Delta_{\mr{Cd}} \neq 0$ and $\Delta_\mr{Bd}=0$. Moreover, we find by varying the ratio between the driving and coupling strength $F/g$, the long-time ergotropy for the latter case behaves as $\frac{\mathcal{E}_\mrb(t\to \infty)}{\omega_\mrb} = F^2/g^2$ agreeing with the result from the resonant case. Finally, we also see from Fig.~\ref{fig:HOfig2app} that even for the case of $\Delta_{\mr{CB}}\neq0$ and $\Delta_\mr{Bd}=0$, for $\vert \Delta_{\mr{CB}}\vert < g$, the steady state ergotropy in the non-zero dephasing case is smaller than the maximum value taken by the oscillating ergotropy in the closed case.

\section{\label{app:C} TLS-HO Model}

Finally, to illustrate the generality of the central results of our paper even further, we consider the scenario of a driven TLS charger connected to an HO battery with the Hamiltonian given by
\begin{align}
\Hop &= \omega_\mrc \hsp_{\mr{C}}\hsm_{\mr{C}}+\omega_\mrb \Bopd\Bop + g(\hsp_{\mr{C}}\Bop + \hsm_\mrc\Bopd)\nonumber \\
& + F(\hsm_\mrc e^{i\omega_{\mr{d}} t} + \hsp_\mrc e^{-i\omega_{\mr{d}} t}) \label{eq:bareHTLSHO}.
\end{align}
The dynamics of interest, as before, is given by the master equation
\begin{align}
     \frac{d\hrho(t)}{dt} &= -i\commu{\Hop}{\hrho(t)} + \frac{\gamma_\mrc}{4} \left(\hsz_\mrc \hrho(t) \hsz_\mrc - \hrho(t)\right)
    \label{eq:TLSmasterhybrid}.
\end{align}
Similar to the two-TLS and two-HO cases, we can derive equations of motion for the operator expectation values. However, for brevity, we numerically simulate the master equation~\eqref{eq:TLSmasterhybrid} \cite{Qutip}, which is nothing but a driven version of the Jaynes-Cummings model. As shown in Figs.~\ref{fig:figHOTLS1} and \ref{fig:figHOTLS2}, we see that for both weak ($F \ll g$) and strong driving ($F \geq g$), there is an optimum dephasing rate at which fastest charging is achieved. Unsurprisingly, even though the battery is given by an HO system, in this case, due to the TLS charger, we have that the optimal dephasing is dependent on the strength of driving.